\documentclass{JHEP3}

\usepackage{graphicx}
\usepackage{amssymb}
\usepackage{amsmath}
\usepackage{latexsym}

\newcommand{\g}{\gamma}
\newcommand{\nslash}{\kern 0.2 em n\kern -0.50em /}
\newcommand{\kslash}{\kern 0.2 em k\kern -0.45em /}
\newcommand{\pslash}{\kern 0.2 em p\kern -0.50em /}
\newcommand{\Sslash}{\kern 0.2 em S\kern -0.50em /}
\newcommand{\Pslash}{\kern 0.2 em P\kern -0.50em /}
\newcommand{\Dslash}{\kern 0.2 em D\kern -0.65em /\kern 0.15em}
\newcommand{\lf}{\left}
\newcommand{\rg}{\right}
\newcommand{\h}{\hat{\bm{h}}}
\newcommand{\eps}{\epsilon}
\newcommand{\ii}{i}                    
\newcommand{\de}{d}                    
\newcommand{\xbj}{x}                   
\newcommand{\zh}{z}
\newcommand{\bm}{\boldsymbol}
\newcommand{\PhiA}{\tilde{\Phi}_{A}}
\newcommand{\DeltaA}{\tilde{\Delta}_{A}}

\newcommand{\slim}{\mskip 1.5mu}              
\newcommand{\Tr}{\operatorname*{Tr}\nolimits} 
\newcommand{\half}{ {\textstyle\frac{1}{2}} }

\newcommand{\cdott}{{\mskip -1.5mu} \cdot {\mskip -1.5mu}}

\allowdisplaybreaks[2]

\title{
Semi-inclusive deep inelastic
scattering at small transverse momentum
}

\author{Alessandro Bacchetta$^{a,\ast}$, Markus Diehl$^{a}$,
Klaus Goeke$^{b}$, Andreas Metz$^{b}$, Piet J. Mulders$^{c}$, 
Marc Schlegel$^{b}$
 \\
$^{a}$Theory Group, Deutsches Elektronen-Synchroton DESY,\\
22603 Hamburg, Germany
\\
$^{b}$Institut f\"ur Theoretische Physik II, Ruhr--Universit\"at Bochum,\\
44780 Bochum, Germany
\\
$^{c}$Dept.\ of Physics and Astronomy, Vrije Universiteit Amsterdam,\\
1081 HV Amsterdam, The Netherlands
\\
$^{\ast}$E-mail: \email{alessandro.bacchetta@desy.de}
}


\abstract
{We study the cross section for one-particle inclusive deep inelastic
scattering off the nucleon for low transverse momentum of the detected
hadron.  We decompose the cross section in terms of structure
functions and calculate them at tree level 
in terms of transverse-momentum-dependent
parton distribution and fragmentation functions.  Our results are
complete in the one-photon exchange approximation at leading and first
subleading twist accuracy, with both beam and target polarization.
}

\preprint{DESY 06-204}

\begin{document}

\bibliographystyle{myJHEP}

\section{Introduction}
\label{sec:intro}

In one-particle-inclusive deep inelastic scattering (DIS) a lepton
scatters off a nucleon and one of the hadrons produced in the
collision is detected.  In the one-photon exchange approximation, the
lepton-nucleon interaction proceeds via a photon of virtuality $Q$.
The cross section depends in particular 
on the azimuthal angle of the final state
hadron about the virtual photon axis, as well as on the azimuthal
angle of the target polarization.  In the kinematic region where the
transverse momentum of the outgoing hadron is low compared to $Q$, the
cross section can be expressed in terms of
transverse-momentum-dependent parton distribution functions (PDFs) and
fragmentation functions (FFs). These partonic functions are
generalizations of the distribution and fragmentation functions
appearing in standard collinear factorization.  They are often
referred to as unintegrated functions, as they are not integrated over
the transverse momentum.

The most complete treatment to date of one-particle-inclusive deep
inelastic scattering at small transverse momentum remains the work of
Mulders and Tangerman~\cite{Mulders:1996dh}, complemented by
Refs.~\cite{Boer:1997nt,Boer:1999uu}.  In the last ten years, however,
the subject has seen important theoretical and experimental progress.
Initiated by the
calculation of a nonvanishing Sivers effect in Ref.~\cite{Brodsky:2002cx}, 
unexpected developments
arose with the correct treatment of Wilson lines in
the definition of transverse-momentum-dependent PDFs and
FFs~\cite{Ji:2002aa,Belitsky:2002sm}.    
In particular, the fundamental tenet of
universality of PDFs and FFs was
revised~\cite{Collins:2002kn,Metz:2002iz,Collins:2004nx}.  New
factorization proofs for the process under consideration here were put
forward~\cite{Ji:2004wu,Ji:2004xq}, updating past work~\cite{Collins:1981uk}.
Some relations proposed in Ref.~\cite{Mulders:1996dh} turned out to be
invalid~\cite{Kundu:2001pk,Goeke:2003az}, and three new PDFs were
discovered~\cite{Bacchetta:2004zf,Goeke:2005hb}.  In the meanwhile,
several experimental measurements of azimuthal asymmetries in
semi-inclusive DIS were
performed~\cite{Breitweg:2000qh,Airapetian:2000tv,Airapetian:2001eg,%
Airapetian:2002mf,Chekanov:2002sz,Airapetian:2004tw,Airapetian:2005jc,%
Alexakhin:2005iw,Avakian:2003pk,Chekanov:2006gt}.

We consider it timely to present in a single, self-contained paper the
results for one-particle-inclusive deep inelastic scattering at small
transverse momentum, in particular including in the cross section all
functions recently introduced.  In Section~\ref{s:struc-fun} we recall
the general form of the cross section for polarized semi-inclusive DIS
and parameterize it in terms of suitable structure functions.  In
Section~\ref{s:tmd-fcts} we give the full parameterization of
quark-quark and quark-gluon-quark correlation functions up to twist three
and review the relations between these functions which are due to the
QCD equations of motion.  The structure functions for semi-inclusive DIS
at small transverse momentum and twist-three accuracy are given in
Section~\ref{s:cross}, and Section~\ref{s:conc} contains our
conclusions.  The relation of the structure functions in the present
paper with the parameterization in Ref.~\cite{Diehl:2005pc} is given
in Appendix~\ref{a:sapeta}, and results for one-jet production in DIS
are listed in Appendix~\ref{a:jet}.

\section{The cross section in terms of structure functions}
\label{s:struc-fun}

We consider the process
\begin{equation}
  \label{sidis}
\ell(l) + N(P) \to \ell(l') + h(P_h) + X ,
\end{equation}
where $\ell$ denotes the beam lepton, $N$ the nucleon target, and $h$
the produced hadron, and where four-momenta are given in parentheses.
Throughout this paper we work in the one-photon exchange approximation
and neglect the lepton mass. We denote by $M$ and $M_h$ the respective
masses of the nucleon and of the hadron $h$.  As usual we define $q =
l - l'$ and $Q^2 = - q^2$ and introduce the variables
\begin{align}
  \label{xyz}
\xbj &= \frac{Q^2}{2\,P\cdott q},
&
y &= \frac{P \cdott q}{P \cdott l},
&
z &= \frac{P \cdott P_h}{P\cdott q},
&
\gamma &= \frac{2 M x}{Q} .
\end{align}
Throughout this section we work in the target rest frame.  Following
the Trento conventions~\cite{Bacchetta:2004jz} we define the azimuthal
angle $\phi_h$ of the outgoing hadron by
\begin{align}
  \label{phi-h-def}
\cos\phi_h &= - \frac{l_\mu P_{h \nu}\, g_\perp^{\mu\nu}}{%
  \sqrt{\slim l_\perp^2\, P_{h \perp}^2}} \,,
&
\sin\phi_h &= - \frac{l_\mu P_{h \nu}\, \epsilon_\perp^{\mu\nu}}{%
  \sqrt{\slim l_\perp^2\, P_{h \perp}^2}} \,,
\end{align}
where $l_\perp^\mu = g_\perp^{\mu\nu} l^{}_\nu$ and $P_{h \perp}^\mu =
g_\perp^{\mu\nu} P^{}_{h \nu}$ are the transverse components of $l$
and $P_h$ with respect to the photon momentum.  The tensors
\begin{align}
g_\perp^{\mu\nu} &= g_{}^{\mu\nu} 
  - \frac{q^\mu P^\nu + P^\mu q^\nu}{P\cdott q\, (1+\gamma^2)}
  + \frac{\gamma^2}{1+\gamma^2} \left(
      \frac{q^\mu q^\nu}{Q^2} - \frac{P^\mu P^\nu}{M^2} \right) ,
\\
\epsilon_\perp^{\mu\nu} &= \epsilon_{}^{\mu\nu\rho\sigma}
   \frac{P_\rho\, q_\sigma}{P\cdott q\, \sqrt{1+\gamma^2}}
\end{align}
have nonzero components $g_\perp^{11} = g_\perp^{22} = -1$ and
$\epsilon_\perp^{12} = - \epsilon_\perp^{21} = 1$ in the coordinate
system of Fig.~\ref{f:angles}, our convention for the totally
antisymmetric tensor being $\epsilon^{0123} = 1$.
We decompose the covariant spin vector $S$ of the target as
\begin{align}
  \label{Sperp-def}
S^\mu &= S_\parallel\, 
   \frac{P^\mu - q^\mu\slim M^2/(P\cdott q)}{M \sqrt{1+\gamma^2}}
   + S^\mu_\perp \,,
&
S_\parallel
  &= \frac{S\cdott q}{P\cdott q}\, \frac{M}{\sqrt{1+\gamma^2}} \,,
&
S^\mu_\perp &= g_\perp^{\mu\nu} S^{}_{\nu}
\end{align}
and define its azimuthal angle $\phi_S$ in analogy to $\phi_h$ in
Eq.~\eqref{phi-h-def}, with $P_h$ replaced by $S$.  Notice that the
sign convention for the longitudinal spin component is such that the
target spin is parallel to the virtual photon momentum for
$S_\parallel = -1$.
The helicity of the lepton beam is denoted by
$\lambda_e$.  We consider the case where the detected hadron $h$ has
spin zero or where its polarization is not measured.

\FIGURE[ht]{
\includegraphics[width=10cm]{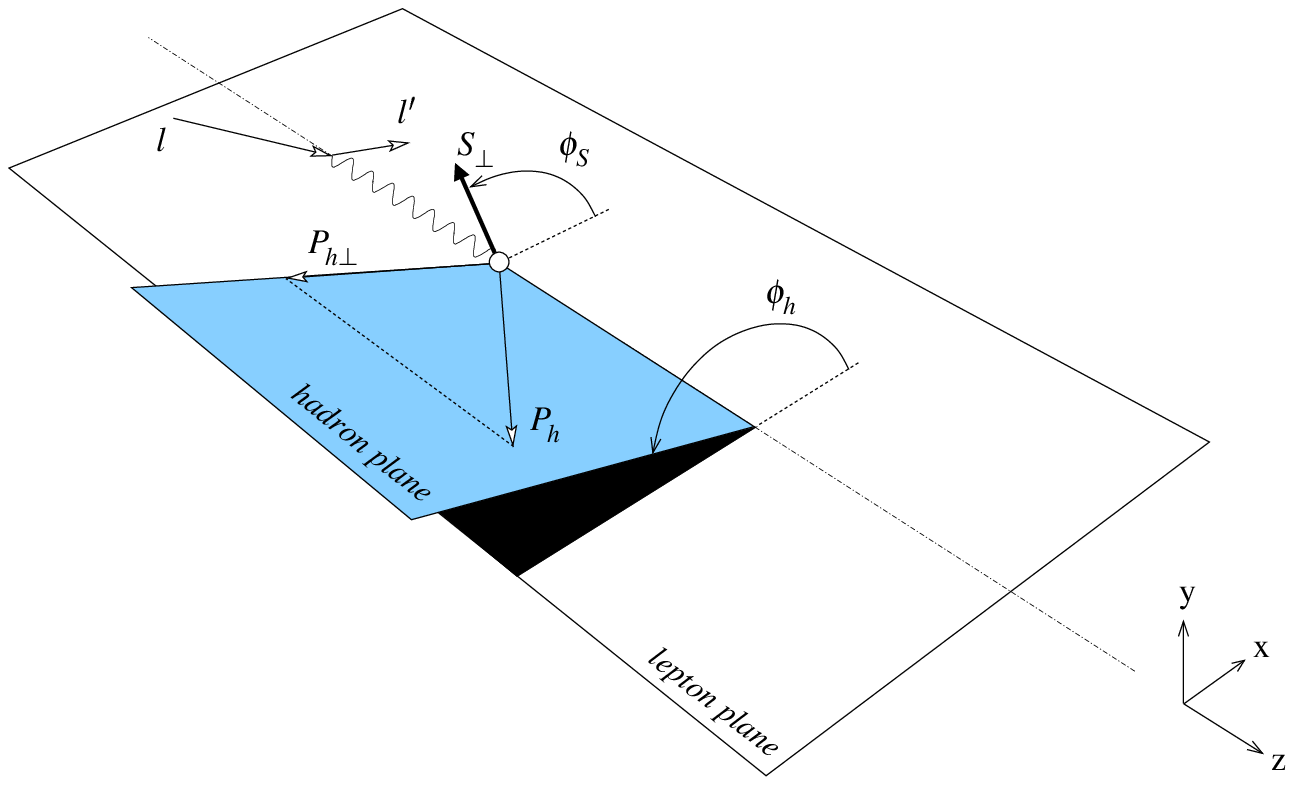}
\caption{\label{f:angles} Definition of azimuthal angles for
  semi-inclusive deep inelastic scattering in the target rest
  frame~\cite{Bacchetta:2004jz}. 
  $P_{h\perp}$ and $S_\perp$ are the transverse parts of $P_h$ and $S$
  with respect to the photon momentum.}
}

Assuming single photon exchange, the lepton-hadron cross section can
be expressed in a model-independent way by a set of structure
functions, see e.g.\
Refs.~\cite{Gourdin:1973qx,Kotzinian:1994dv,Diehl:2005pc}.  We 
use here a modified version of the notation in Ref.~\cite{Diehl:2005pc},
see App.~\ref{a:sapeta}, and write\footnote{%
The polarizations $S_L$ and $S_T$ in \protect\cite{Diehl:2005pc} have
been renamed to $S_\parallel$ and $|\bm{S}_\perp|$ here.  This is to
avoid a clash of notation with Section~\protect\ref{s:tmd-fcts}, where
subscripts $L$ and $T$ refer to a different $z$-axis than in
Fig.~\protect\ref{f:angles}.}
\begin{align}
\lefteqn{\frac{d\sigma}{d\xbj \, dy\, d\psi \,dz\, d\phi_h\, d P_{h\perp}^2}
=}
\nonumber \\ & \quad 
\frac{\alpha^2}{\xbj y\slim Q^2}\,
\frac{y^2}{2\,(1-\varepsilon)}\,  \biggl( 1+\frac{\gamma^2}{2\xbj} \biggr)\,
\Biggl\{
F_{UU ,T}
+ 
\varepsilon\slim
F_{UU ,L}
+ \sqrt{2\,\varepsilon (1+\varepsilon)}\,\cos\phi_h\,
F_{UU}^{\cos\phi_h}
\nonumber \\  & \quad \qquad
+ \varepsilon \cos(2\phi_h)\, 
F_{UU}^{\cos 2\phi_h}
+ \lambda_e\, \sqrt{2\,\varepsilon (1-\varepsilon)}\, 
           \sin\phi_h\, 
F_{LU}^{\sin\phi_h}
\phantom{\Bigg[ \Bigg] }
\nonumber \\  & \quad \qquad
+ S_\parallel\, \Bigg[ 
 \sqrt{2\, \varepsilon (1+\varepsilon)}\,
  \sin\phi_h\, 
F_{UL}^{\sin\phi_h}
+  \varepsilon \sin(2\phi_h)\, 
F_{UL}^{\sin 2\phi_h}
\Bigg]
\nonumber \\  & \quad \qquad
+ S_\parallel \lambda_e\, \Bigg[ \,
  \sqrt{1-\varepsilon^2}\; 
F_{LL}
+\sqrt{2\,\varepsilon (1-\varepsilon)}\,
  \cos\phi_h\, 
F_{LL}^{\cos \phi_h}
\Bigg]
\nonumber \\  & \quad \qquad
+ |\bm{S}_\perp|\, \Bigg[
  \sin(\phi_h-\phi_S)\,
\Bigl(F_{UT ,T}^{\sin\lf(\phi_h -\phi_S\rg)}
+ \varepsilon\, F_{UT ,L}^{\sin\lf(\phi_h -\phi_S\rg)}\Bigr)
\nonumber \\  & \quad  \qquad \qquad
+ \varepsilon\, \sin(\phi_h+\phi_S)\, 
F_{UT}^{\sin\lf(\phi_h +\phi_S\rg)}
+ \varepsilon\, \sin(3\phi_h-\phi_S)\,
F_{UT}^{\sin\lf(3\phi_h -\phi_S\rg)}
\phantom{\Bigg[ \Bigg] }
\nonumber \\  & \quad \qquad \qquad
+ \sqrt{2\,\varepsilon (1+\varepsilon)}\, 
  \sin\phi_S\, 
F_{UT}^{\sin \phi_S }
+ \sqrt{2\,\varepsilon (1+\varepsilon)}\, 
  \sin(2\phi_h-\phi_S)\,  
F_{UT}^{\sin\lf(2\phi_h -\phi_S\rg)}
\Bigg]
\nonumber \\  & \quad \qquad 
+ |\bm{S}_\perp| \lambda_e\, \Bigg[
  \sqrt{1-\varepsilon^2}\, \cos(\phi_h-\phi_S)\, 
F_{LT}^{\cos(\phi_h -\phi_S)}
+\sqrt{2\,\varepsilon (1-\varepsilon)}\, 
  \cos\phi_S\, 
F_{LT}^{\cos \phi_S}
\nonumber \\  & \quad \qquad \qquad
+\sqrt{2\,\varepsilon (1-\varepsilon)}\, 
  \cos(2\phi_h-\phi_S)\,  
F_{LT}^{\cos(2\phi_h - \phi_S)}
\Bigg] \Biggr\},
\label{e:crossmaster}
\end{align}
where $\alpha$ is the fine structure constant and the structure
functions on the r.h.s.\ depend on $\xbj$, $Q^2$, $z$ and
$P_{h\perp}^2$.  The angle $\psi$ is the azimuthal angle of $\ell'$
around the lepton beam axis with respect to an arbitrary fixed
direction, which in case of a transversely polarized target we choose
to be the direction of $S$.  The corresponding relation between $\psi$
and $\phi_S$ is given in Ref.~\cite{Diehl:2005pc}; in deep inelastic
kinematics one has $\de \psi \approx \de \phi_S$.  The first and
second subscript of the above structure functions indicate the
respective polarization of beam and target, whereas the third
subscript in $F_{UU,T}^{}$, $F_{UU,L}^{}$ and
$F_{UT ,T}^{\sin\lf(\phi_h -\phi_S\rg)}$,
$F_{UT ,L}^{\sin\lf(\phi_h-\phi_S\rg)}$ 
specifies the polarization of
the virtual photon.  Note that longitudinal or transverse target
polarization refer to the photon direction here.  The conversion to
the experimentally relevant longitudinal or transverse polarization
w.r.t.\ the lepton beam direction is straightforward and given in
\cite{Diehl:2005pc}.  The ratio $\varepsilon$ of longitudinal and
transverse photon flux in 
\eqref{e:crossmaster} is given by
\begin{align}
\varepsilon &= \frac{1-y -\frac{1}{4}\slim \gamma^2 y^2}{1-y
  +\frac{1}{2}\slim y^2 +\frac{1}{4}\slim \gamma^2 y^2} ,
\end{align}  
so that the depolarization factors can be written as
\begin{alignat}{2}
\frac{y^2}{2\,(1-\varepsilon)} &= 
\frac{1}{1+\gamma^2} \lf(1-y+\tfrac{1}{2}\slim y^2 
  + \tfrac{1}{4}\slim \gamma^2 y^2 \rg)
&&\approx \left(1-y +\tfrac{1}{2}\slim y^2\right),
\\
\frac{y^2}{2\,(1-\varepsilon)}\,\varepsilon &= 
\frac{1}{1+\gamma^2} \lf(1-y- \tfrac{1}{4}\slim \gamma^2 y^2 \rg) 
&&\approx (1-y) ,
\\
\frac{y^2}{2\,(1-\varepsilon)}\,\sqrt{2\,\varepsilon(1+\varepsilon)} &= 
\frac{1}{1+\gamma^2}\;(2-y) \,
   \sqrt{1-y - \tfrac{1}{4}\slim \gamma^2 y^2}
&&\approx (2-y)\, \sqrt{1-y}, 
\phantom{\frac{1}{\sqrt{\gamma^2}}}
\\
\frac{y^2}{2\,(1-\varepsilon)}\,\sqrt{2\,\varepsilon(1-\varepsilon)} &=
\frac{1}{\sqrt{1+\gamma^2}}\;y \,
   \sqrt{1-y - \tfrac{1}{4}\slim \gamma^2 y^2}
&&\approx y \,  \sqrt{1-y},
\\
\frac{y^2}{2\,(1-\varepsilon)}\,\sqrt{1-\varepsilon^2} &= 
\frac{1}{\sqrt{1+\gamma^2}}\; y \lf(1-\tfrac{1}{2}\slim y \rg) 
&&\approx y\left(1 -\tfrac{1}{2}\slim y\right).
\end{alignat} 
Integration of Eq.~(\ref{e:crossmaster}) over the transverse momentum
$\bm{P}_{h\perp}$ of the 
outgoing hadron gives the semi-inclusive deep inelastic
scattering cross section
\begin{multline}
\frac{d\sigma}{d\xbj \, dy\, d\psi \,dz}
=
\frac{2 \alpha^2}{\xbj y\slim Q^2}\,
\frac{y^2}{2\,(1-\varepsilon)}\, \biggl( 1+\frac{\gamma^2}{2\xbj} \biggr)\,
\biggl\{
F_{UU ,T} + \varepsilon\slim F_{UU ,L}
+ S_\parallel \lambda_e\,
  \sqrt{1-\varepsilon^2}\; 
F_{LL}
\\  
+ |\bm{S}_\perp|\,
\sqrt{2\,\varepsilon (1+\varepsilon)}\,
  \sin\phi_S\, 
F_{UT}^{\sin \phi_S }
+ |\bm{S}_\perp| \lambda_e\, \sqrt{2\,\varepsilon (1-\varepsilon)}\, 
  \cos\phi_S\, 
F_{LT}^{\cos \phi_S}
 \biggr\},
\label{e:crossintsidis}
\end{multline}
where the structure functions on the r.h.s.\ are integrated
versions of the previous ones, i.e.\
\begin{equation}
F_{UU ,T}(\xbj,\zh,Q^2) =
  \int \de^2 \bm{P}_{h \perp}\,F_{UU ,T}(\xbj,\zh,P_{h \perp}^2,Q^2)
\end{equation} 
and similarly for the other functions.

We can finally connect the semi-inclusive structure functions to those
for inclusive deep inelastic scattering.  With the energies $E_h = (P
\cdott P_h) /M$ and $\nu = (P \cdott q) /M$ of the detected hadron and
the virtual photon, we have
\begin{equation}
\sum_h \int \de z\, z\,
  \frac{d\sigma(\ell\slim N \to \ell\slim h\slim X)}{dz\,
    d\xbj \, dy\, d\psi}
= \frac{1}{\nu} \sum_h \int \de E_h\, E_h
  \frac{d\sigma(\ell\slim N \to \ell\slim h\slim X)}{dE_h\, 
    d\xbj \, dy\, d\psi}
= \frac{\nu + M}{\nu} \,
  \frac{d\sigma(\ell\slim N \to \ell\slim X)}{d\xbj \, dy\, d\psi} ,
\label{e:sidistodis}
\end{equation}
where we have summed over all hadrons in the final state, whose total
energy is $\nu+M$.  Using $(\nu + M)/\nu = 1 + \gamma^2/(2x)$ we have
\begin{align}
\frac{d\sigma}{d\xbj \, dy\, d\psi}
=
\frac{2 \alpha^2}{\xbj y\slim Q^2}\, \frac{y^2}{2\,(1-\varepsilon)}\, 
&\biggl\{
F_T + \varepsilon\slim F_L
+ S_\parallel \lambda_e\,
  \sqrt{1-\varepsilon^2}\; 
2 \xbj\,(g_1 - \gamma^2 g_2)
\nonumber \\  & \quad
- |\bm{S}_\perp| \lambda_e\, \sqrt{2\,\varepsilon (1-\varepsilon)}\, 
  \cos\phi_S\, 
2 \xbj \gamma\, (g_1 + g_2)
\biggr\},
\end{align} 
where
\begin{alignat}{2}
  \label{dis-fun}
\sum_h \int \de z \,z\, F_{UU ,T}(\xbj,\zh,Q^2)
&= 2\xbj F_1(\xbj,Q^2) 
& &= F_T(\xbj,Q^2) ,
\\
\sum_h \int \de z \,z\, F_{UU ,L}(\xbj,\zh,Q^2)
&= (1+\gamma^2)F_2(\xbj,Q^2) - 2\xbj F_1(\xbj,Q^2) 
& &= F_L(\xbj,Q^2) ,
\\
\sum_h \int \de z \,z\, F_{LL}(\xbj,\zh,Q^2)
&= 2\xbj\,\bigl(g_1(\xbj,Q^2) - \gamma^2 g_2(\xbj,Q^2)\bigr) ,
\\
\sum_h \int \de z \,z\,  F_{LT}^{\cos \phi_S}(\xbj,\zh,Q^2)
&= - 2\xbj \gamma\, \bigl(g_1(\xbj,Q^2) + g_2(\xbj,Q^2)\bigr) 
\label{dis-last}
\end{alignat} 
in terms of the conventional deep inelastic structure functions.  For
the relation with the more common expression for target 
polarization along or
transverse to the lepton beam
direction
see Refs.~\cite{Adams:1994id,Lampe:1998eu,Diehl:2005pc}.  
Finally, time-reversal
invariance requires (see, e.g., Ref.~\cite{Diehl:2005pc})
\begin{equation} 
\sum_h \int \de z \,z\,  F_{UT}^{\sin \phi_S}(\xbj,\zh,Q^2) = 0.
\label{e:nosinphis}
\end{equation}

\section{Transverse-momentum dependent distribution and fragmentation
  functions} 
\label{s:tmd-fcts}

\subsection{Light-cone coordinates}
\label{s:light-cone}

Manipulations with parton distribution and fragmentation functions are
conveniently done using light-cone coordinates.  For an arbitrary
four-vector $v$ we write $v^\pm = (v^0 \pm v^3) /\sqrt{2}$ and $\bm{v}_T =
(v^1, v^2)$ in a specified reference frame 
and give all components as $[v^-,v^+,\bm{v}_T]$.  We will
use the transverse tensors $g_T^{\alpha\beta}$ and
$\epsilon_T^{\alpha\beta}$, whose only nonzero components are
$g_T^{11} = g_T^{22} = -1$ and $\epsilon_T^{12} = - \epsilon_T^{21} =
1$.  The light-cone decomposition of a vector can be written in a
Lorentz covariant fashion using two light-like vectors $n_+ =
[0,1,\bm{0}_T]$ and $n_- = [1,0,\bm{0}_T]$ and
promoting $\bm{v}_T$ to a four-vector $v_T = [0,0,\bm{v}_T]$.  One then
has
\begin{equation}
  v^\mu  = v^+ n_+^\mu + v^- n_-^\mu + v_T^\mu ,
\end{equation}
where $v^+ = v \cdott n_-$, $v^- = v \cdott n_+$ and $v_T \cdott n_+ =
v_T \cdott n_- = 0$.  We further have
\begin{align}
  \label{e:eps-def}
g_T^{\alpha\beta} &= g^{\alpha\beta} - n_{+}^\alpha n_{-}^\beta -
n_{-}^\alpha n_{+}^\beta ,
&
\epsilon_T^{\alpha\beta} &= \epsilon^{\alpha\beta\rho\sigma}\, 
                             n_{+\rho} n_{-\sigma} .
\end{align}
Note that scalar products with transverse
four-vectors are in Minkowski space, so that $v_T \cdott w_T = -
\bm{v}_T \cdott \bm{w}_T$.

For the discussion of distribution functions we will choose light-cone
coordinates such that $P$ has no transverse component, i.e.
\begin{equation}
  \label{P-frame}
P^\mu = P^+ n_+^\mu + \frac{M^2}{2P^+}\, n_-^\mu \,.
\end{equation}
The spin vector of the target can then be decomposed as
\begin{equation}
  \label{ST-def}
S^\mu = S_L\, 
  \frac{(P\cdott n_-)\, n_+^\mu - (P\cdott n_+)\, n_-^\mu}{M} + S_T^\mu \,,
\end{equation}
which implies $S_L = M\slim (S\cdott n_-) /(P\cdott n_-)$.  
Similarly, for the discussion of
fragmentation functions, we will assume a coordinate choice with
\begin{equation}
  \label{h-frame}
P_h^{\mu} = P_h^- n_-^\mu + \frac{M_h^2}{2P_h^-}\, n_+^\mu \,.
\end{equation}

\subsection{Calculation of the hadronic tensor}
\label{s:hadtens}

We consider semi-inclusive DIS in the kinematical limit where $Q^2$
becomes large while $x$, $z$ and $P_{h\perp}^2$ remain fixed, and will
perform an expansion in powers of $1/Q$.
For the calculation we use a frame where both
\eqref{P-frame} and \eqref{h-frame} are satisfied, and where $x P^+ =
P_h^- /z = Q/\sqrt{2}$.  Notice that this differs from the choice in
Section~\ref{s:struc-fun}, where the transverse direction was defined
with respect to the momenta of the target and the virtual photon,
instead of the momenta of the target and the produced hadron.  Details
on the relation between the two choices can be found in
\cite{Mulders:1996dh,Boer:2003cm}.  In particular, $S_L$ and $S_T$
defined by \eqref{ST-def} with \eqref{P-frame} and \eqref{h-frame}
differ from $S_\parallel$ and $S_\perp$ in \eqref{Sperp-def} by terms
of order $1/Q^2$ and $1/Q$, respectively.

The leptoproduction cross section can be expressed as the contraction
of a hadronic and a leptonic tensor,
\begin{equation}
\frac{\de \sigma}{\de\xbj\,\de y\,
\de \psi\, \de z\,\de\phi_h\, \de P_{h \perp}^2} 
= \frac{\alpha^2\slim y}{8 z\slim Q^4}\, 2 M W^{\mu \nu}\,
  L_{\mu \nu} \, , 
\label{eq:cross}
\end{equation}
where the leptonic tensor is given by
\begin{equation}
 L_{\mu \nu} =
2 \lf(l^{}_{\mu}\slim l\slim'_{\nu} + l\slim'_{\mu}\slim l^{}_{\nu} 
  - l\cdott l\slim'\slim g_{\mu \nu}^{} \rg) 
+ 2 \ii \lambda_e\,
\eps_{\mu \nu \rho \sigma}\, l^{\rho} l'^{\sigma}.
\end{equation} 
The hadronic tensor is defined as
\begin{equation}
2M W^{\mu\nu} = \frac{1}{(2 \pi)^3} \sum_{X}  
\int \frac{\de^3 \! \bm{P}_{X}} {2 P_{X}^0}
  \;\delta^{(4)} \bigl( q+P-P_{X}-P_h \bigr)\,
\langle P| J^{\mu} (0) |\slim h, {X}\rangle \langle h, {X}| 
   J^{\nu} (0)|P\rangle ,
\end{equation}
where $J^\mu(\xi)$ is the electromagnetic current divided by the
elementary charge and a sum is implied over the polarizations of all
hadrons in the final state.

\FIGURE[ht]{
\begin{tabular}{ccc}
\includegraphics[height=4.5cm]{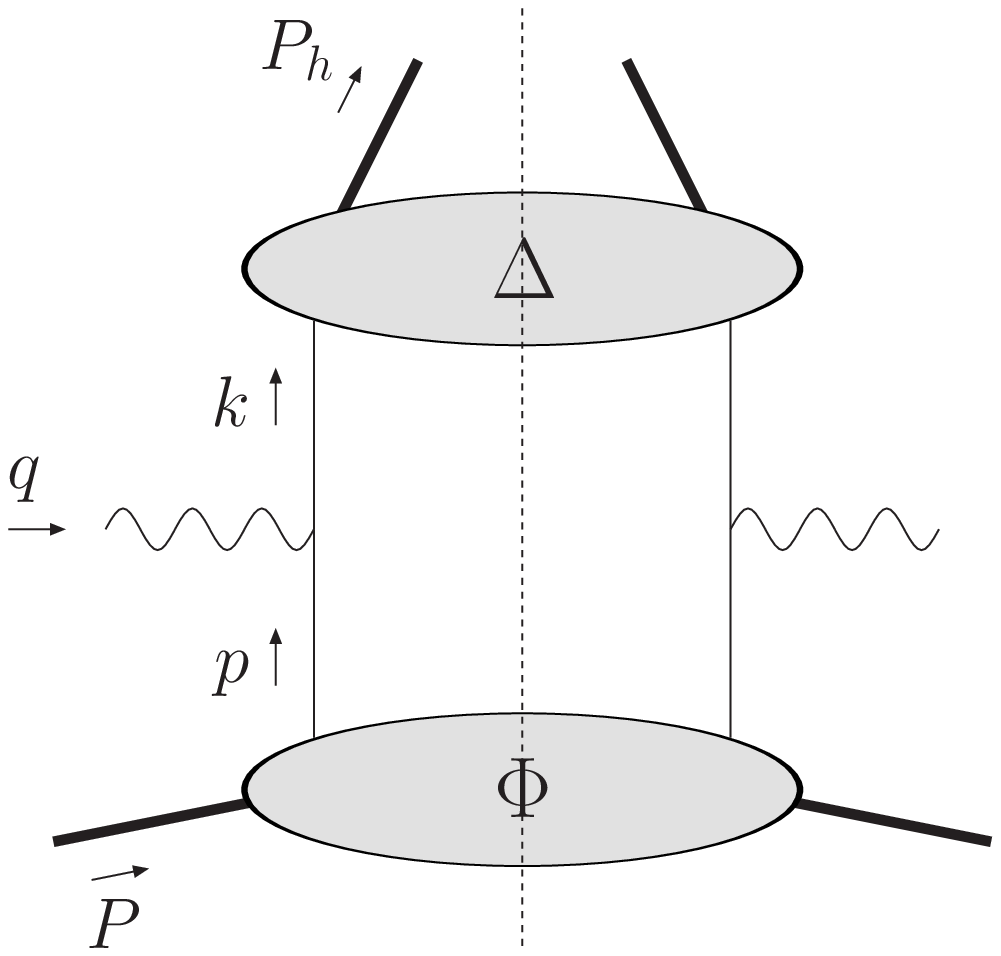}
&&
\includegraphics[height=4.5cm]{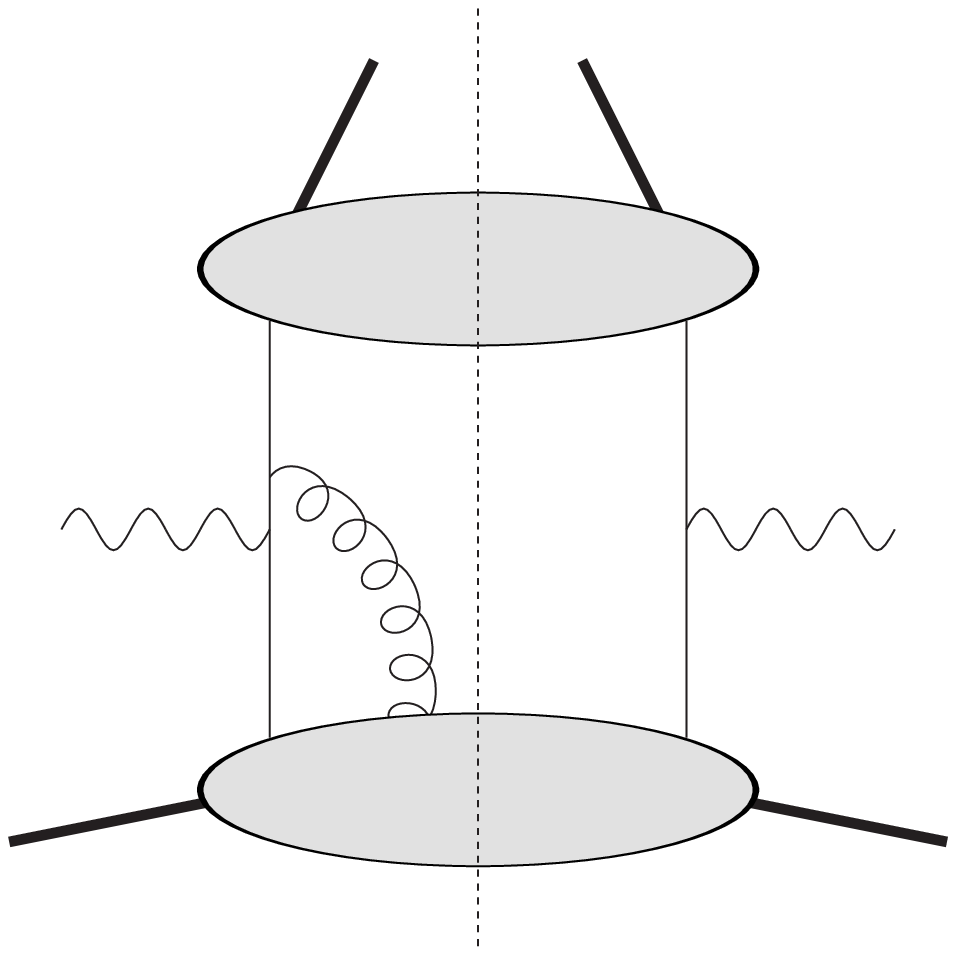}
\\
(a)
&&
(b)
\\
\\
\includegraphics[height=4.5cm]{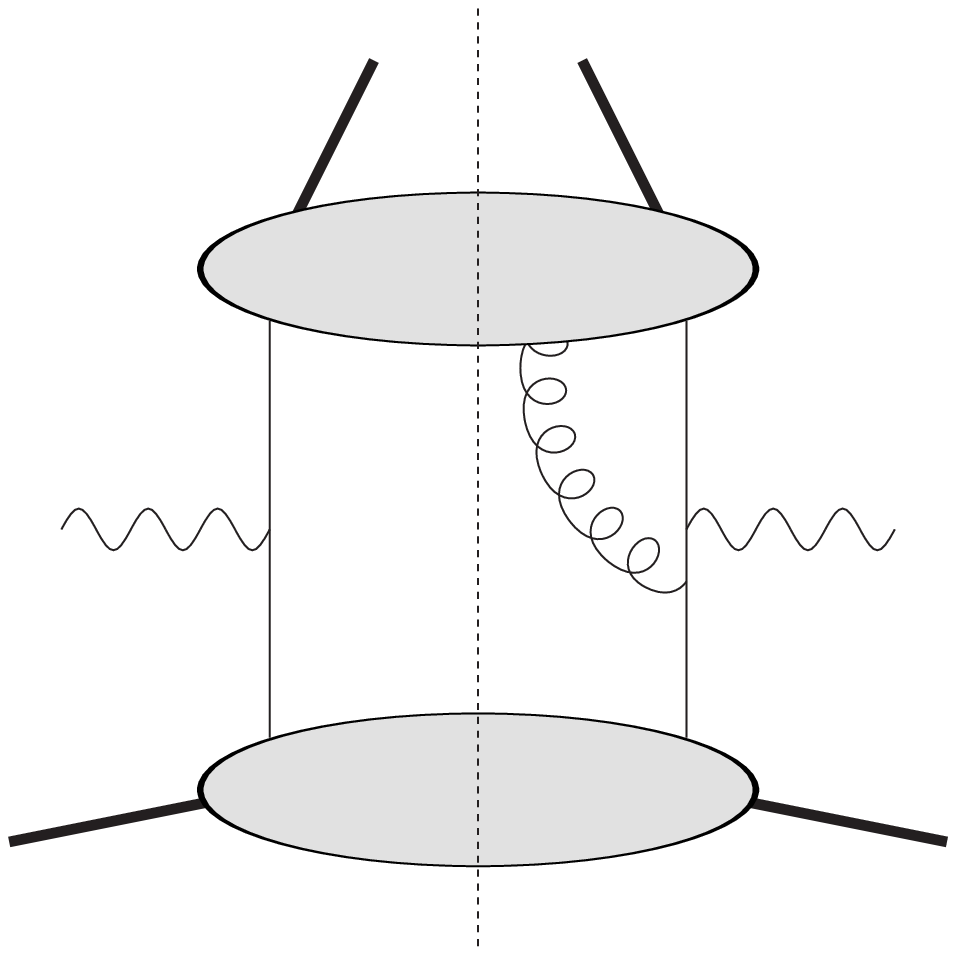}
&&
\includegraphics[height=4.5cm]{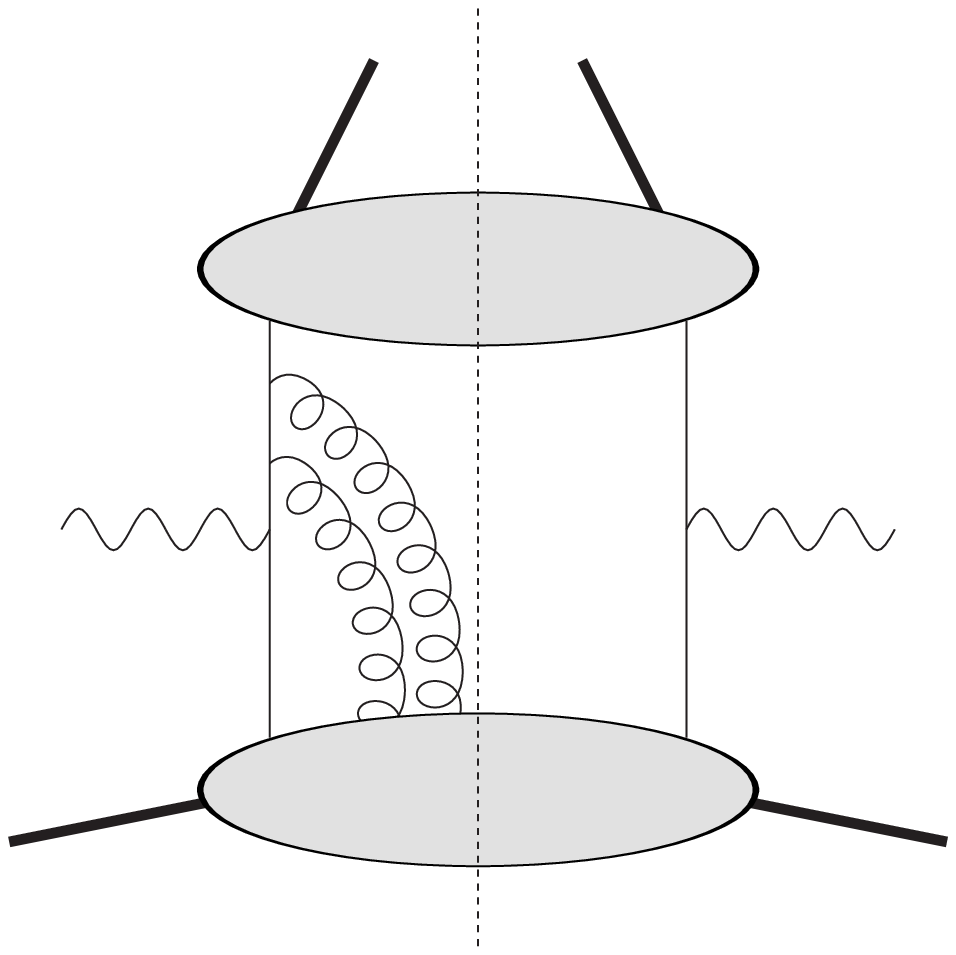}
\\
(c)
&&
(d)
\end{tabular}
\caption{\label{f:graphs}  Examples of graphs
  contributing to semi-inclusive DIS at low transverse momentum of the
  produced hadron.}
}

The calculations in this paper are based on the factorization of the
cross section into a hard photon-quark scattering process and nonperturbative
functions describing the distribution of quarks in the target or the
fragmentation of a quark into the observed hadron.  We limit ourselves
to the leading and first subleading term in the $1/Q$ expansion of the
cross section and to graphs with the hard scattering at tree level.
Loops can then only occur as shown in Fig.~\ref{f:graphs}b, c, d, with
gluons as external legs of the nonperturbative functions.  The corresponding
expression of the hadronic tensor is~\cite{Mulders:1996dh,Boer:2003cm}
\begin{multline}
2M W^{\mu\nu} 
= 2z \sum_a   e_a^2 \; \int
\de^2 \bm{p}_T\ \de^2 \bm{k}_T^{}\ 
\delta^2(\bm{p}_T + \bm{q}_T - \bm{k}_T^{}) \;
\Tr \biggl\{
  \Phi^a (x,p_T) \gamma^\mu \Delta^a (z,k_T) \gamma^\nu
\\
- \frac{1}{Q\sqrt{2}} \biggl[
  \gamma^\alpha \nslash_+ \gamma^\nu \slim
  \tilde{\Phi}^a_{A\slim \alpha}(x,p_T)\slim \gamma^\mu \Delta^a(z,k_T) 
+ \gamma^\alpha \nslash_- \gamma^\mu
  \tilde{\Delta}^a_{A\slim \alpha} (z,k_T)\slim \gamma^\nu 
  \Phi^a (x,p_T)  
  + \mathrm{h.c.} \slim\biggr] \biggr\},\\[0.2em]
\label{eq1}
\end{multline}
with corrections of order $1/Q^2$, where the sum runs over the quark
and antiquark flavors $a$, and $e_a$ denotes the fractional charge of
the struck quark or antiquark.  In the next subsections we discuss in
detail the correlation functions $\Phi$ for quark distributions, $\Delta$ for
quark fragmentation, and their analogs $\PhiA$ and $\DeltaA$ with an
additional gluon leg.  The first, second and third term in
Eq.~\eqref{eq1} respectively correspond to the graphs in
Fig.~\ref{f:graphs}a, b and c, with gluons having transverse
polarization.  The analogs of Fig.~\ref{f:graphs}b and c with the
gluon on the other side of the final-state cut correspond to the
``h.c.'' terms in Eq.~\eqref{eq1}.

The Wilson lines needed in the color gauge invariant soft functions come
from graphs with additional gluons exchanged between the hard scattering
and either the distribution or the fragmentation function (as in
Fig.~\ref{f:graphs}b,c,d).  The corresponding gluons have
polarization vectors proportional to $n_+$ in the first and to $n_-$
in the second case, except for contributions from the gluon potential
at infinity.  The importance of the latter has been discussed in
\cite{Ji:2002aa,Belitsky:2002sm}, and a detailed derivation of the
Wilson lines appearing in semi-inclusive DIS was given in
Ref.~\cite{Boer:2003cm} for the leading terms in the $1/Q$ expansion.
For the contributions subleading in $1/Q$ only the cross section
integrated over $\bm{P}_{h \perp}$ has been analyzed in the same
reference.  

Going beyond the tree graphs just discussed
requires modifications in the factorization formula
\eqref{eq1}. In particular, radiative corrections involving low-frequency 
gluons 
 introduce Sudakov
logarithms and so-called soft factors.  
A proof of factorization to all orders in $\alpha_s$ for
the (similar but simpler) case of two-hadron production in $e^+e^-$
collisions was given long ago \cite{Collins:1981uk}.  Recent work on
all-order factorization in semi-inclusive DIS can be found in
Refs.~\cite{Ji:2004wu,Ji:2004xq} and in Ref.~\cite{Collins:2004nx}.
Whether and how the tree-level factorization used in the present paper
extends to subleading level in $1/Q$ is presently not known.

\subsection{The quark-quark correlators}
\label{s:qqcorr}

The quark-quark distribution correlation function is defined
as\footnote{%
To be precise, one should distinguish the momentum fraction $x$ in the
definition of distribution functions from the Bjorken variable defined
in Eq.~\protect\eqref{xyz}.  They coincide however in the process we
consider, so that we drop this distinction for simplicity.  An
analogous remark holds for the argument $z$ in the fragmentation
functions below.}
\begin{equation}  
{\Phi_{ij}} 
(x,p_T)= \int 
        \frac{\de \xi^- \de^2 \bm{\xi}_T}{(2\pi)^{3}}\; 
 e^{\ii p \cdot \xi}\,
       \langle P|\bar{\psi}_j(0)\,
{\cal U}^{n_-}_{(0,+\infty)}\,
{\cal U}^{n_-}_{(+\infty,\xi)}\,
\psi_i(\xi)|P \rangle \bigg|_{\xi^+=0}
\label{e:phi} 
 \end{equation}   
with $p^+ = x P^+$, where here and in the following we omit the flavor
index $a$.  The corresponding correlator for antiquarks is obtained by
replacing the quark field by its transform under charge conjugation,
see Ref.~\cite{Mulders:1996dh}.  In the correlator \eqref{e:phi} we have
gauge links (Wilson lines)
\begin{align} 
{\cal U}^{n_-}_{(0,+\infty)} &=
   {\cal U}^{n_-}{(0^-, \infty^-;\bm{0}_T)}\;
   {\cal U}^T{(\bm{0}_T, \bm{\infty}_T;\infty^-)},
\label{e:wilson1}
\\[0.2em]
{\cal U}^{n_-}_{(+\infty,\xi)} &=
   {\cal U}^T{(\bm{\infty}_T, \bm{\xi}_T;\infty^-)}\;
   {\cal U}^{n_-}{(\infty^-, \xi^-,\bm{\xi}_T)}.
\label{e:wilson2}
\end{align}  
Here ${\cal U}^{n_-}{(a^-,b^-;\bm{c}_T)}$ indicates a Wilson line
running along the minus direction from $[a^-,0^+,\bm{c}_T]$ to
$[b^-,0^+,\bm{c}_T]$, while ${\cal U}^T{(\bm{a}_T,\bm{b}_T;c^-)}$
indicates a Wilson line running in the transverse direction from
$[c^-,0^+,\bm{a}_T]$ to $[c^-,0^+,\bm{b}_T]$, i.e.\
\begin{align} 
{\cal U}^{n_-}{(a^-,b^-;\bm{c}_T)} &= 
{\cal P} \exp \biggl[ -\ii g \int_{a^-}^{b^-} 
  \de \zeta^- A^+ (\zeta^-,0^+,\bm{c}_T)  \biggr] ,
\\
{\cal U}^T{(\bm{a}_T,\bm{b}_T;c^-)} &= 
{\cal P} \exp \biggl[ -\ii g \int_{\bm{a}_T}^{\bm{b}_T} 
  \de {\zeta}_T\cdott {A}_T (c^-,0^+,\bm{\zeta}_T)  \biggr] .
\end{align} 
The correlator in Eq.~(\ref{e:phi}) is the one appearing in
semi-inclusive DIS.  In different processes the structure of the gauge
link can
change~\cite{Collins:2002kn,Bomhof:2004aw,Bacchetta:2005rm,Bomhof:2006dp}.
For instance, in Drell-Yan lepton pair production all occurrences of
$\infty^-$ in the gauge links should be replaced by $-\infty^-$.  In
particular, this reverses the sign of all T-odd distribution functions
appearing in the correlator (see below).  In partonic processes with
colored states in both the initial and final state, the gauge link
contains contributions running to $\infty^-$ as well as $-\infty^-$,
and T-odd terms differ by more than a simple sign change.
Alternatively, it is possible to work always with the correlator for
semi-inclusive DIS when convoluting T-odd functions with so-called
gluonic-pole cross sections instead of the normal partonic cross
sections \cite{Bacchetta:2005rm,Bomhof:2006ra}.  We note that beyond
tree-level the Wilson lines in Eqs.~(\ref{e:wilson1}) and
(\ref{e:wilson2}) lead to logarithmic divergences in the correlator
\eqref{e:phi}.  These are due to gluons with vanishing momentum
component along $n_+$ and need to be regularized.  The twist-two part
of the correlator can be regularized in ways consistent with
factorization to leading power in
$1/Q$~\cite{Collins:1981uk,Collins:2003fm,Ji:2004wu,Collins:2004nx}.
It is presently not known how to extend this to the twist-three
sector, see Ref.~\cite{Gamberg:2006ru} for an investigation of this
case.

A complete parameterization of the quark-quark correlation function
has been given in Ref.~\cite{Goeke:2005hb}.  Here we limit
ourselves to the twist-three level, where we have
\begin{align}
\Phi(x,p_T) &= \frac{1}{2}\, \biggl\{ 
f_1 \nslash_+ 
- {f_{1T}^\perp}\, \frac{\eps_T^{\rho \sigma} p_{T\rho}^{}\slim
  S_{T\sigma}^{}}{M} \, \nslash_+ 
+ g_{1s} \gamma_5\nslash_+
\nonumber \\[0.2em] & \quad \qquad
+h_{1T}\,\frac{\bigl[\Sslash_T, \nslash_+ \bigr]\gamma_5}{2}
+ h_{1s}^\perp \,\frac{\bigl[\pslash_T, \nslash_+ \bigr]\gamma_5}{2 M}
+\ii \, {h_1^\perp} \frac{ \bigl[\pslash_T, \nslash_+ \bigr]}{2M}
\biggr\} 
\nonumber \\[0.2em] & \quad
+ \frac{M}{2 P^+}\,\biggl\{ 
e 
- \ii\,{e_s} \,\gamma_5 
- {e_{T}^\perp}\, \frac{\eps_T^{\rho \sigma} p_{T\rho}^{}\slim
  S_{T\sigma}^{}}{M} 
\nonumber \\ & \quad \qquad \qquad
+ f^\perp\, \frac{\pslash_T}{M}
- {f_T'}\,\epsilon_T^{\rho\sigma} \gamma_\rho^{}\slim S_{T \sigma}^{}
- {f_s^{\perp}}\,\frac{\eps_T^{\rho \sigma}
    \g_{\rho}^{}\slim p_{T \sigma}^{}}{M}
\nonumber \\ & \quad \qquad \qquad
+ g_T'\, \gamma_5\Sslash_T
+ g_s^{\perp} \gamma_5 \frac{\pslash_T}{M}
- {g^\perp} \g_5\,\frac{\eps_T^{\rho \sigma} 
    \g_{\rho}^{}\slim p_{T \sigma}^{}}{M}
\nonumber \\[0.2em] & \quad \qquad  \qquad
+ h_s\,\frac{[\nslash_+, \nslash_-]\gamma_5}{2}
+ h_T^{\perp}\,\frac{\bigl[\Sslash_T, \pslash_T \bigr]\gamma_5}{2 M}
+ \ii \, {h} \frac{ \bigl[\nslash_+, \nslash_- \bigr]}{2} 
\biggr\} .
\label{eq:phi}
\end{align}
The distribution functions on the r.h.s.\ depend on $x$ and $p_T^2$,
except for the functions with subscript $s$, where we use the
shorthand notation~\cite{Mulders:1996dh}
\begin{equation} 
g_{1s}(x, p_T) 
= S_L\,g_{1L}(x, p_T^2) - \frac{p_T \cdott S_T}{M}\,g_{1T}(x, p_T^2)
\label{eq:shorthand}
\end{equation} 
and so forth for the other functions.
The first eight distributions of Eq.~\eqref{eq:phi} are referred to as twist
two, and the next 
16 distributions are referred to as twist three,
where we use the notion of ``dynamical twist'' as explained in
Ref.~\cite{Jaffe:1996zw}. 
  The remaining eight functions of twist
four have been omitted here and can be found in Ref.~\cite{Goeke:2005hb}.
The 10 functions $f_{1T}^{\perp}$, $h_1^{\perp}$, $e_L^{}$, $e_T^{}$,
$e_T^{\perp}$, $f_T'$, $f_L^{\perp}$, $f_T^{\perp}$, $g^{\perp}$, $h$
are T-odd~\cite{Boer:1997nt,Goeke:2005hb}, i.e.\ they change sign
under ``naive time reversal'', which is defined as usual time
reversal, but without the interchange of initial and final states.
The functions $g^{\perp}$~\cite{Bacchetta:2004zf}, $e_T^{\perp}$ and
$f_T^{\perp}$~\cite{Goeke:2005hb} exist because the direction of the
Wilson line provides a vector independent of $P$ and $S$ for a Lorentz
invariant decomposition of the correlation function $\Phi(p)$, which
is defined as in \eqref{e:phi} but with $\xi^+$ integrated over
instead of being set to zero~\cite{Goeke:2003az}.  The notation used
in \eqref{eq:phi} is consistent with
Refs.~\cite{Mulders:1996dh,Boer:1997nt,Bacchetta:2004zf,Goeke:2005hb},
except for three points of discrepancy: $(i)$ the sign of $g^\perp$ is
opposite to that in Ref.~\cite{Bacchetta:2004zf}, where the function
was originally introduced, and consistent with
Ref.~\cite{Goeke:2005hb}; $(ii)$ the sign of the function
$f_{L}^{\perp}$ is opposite to that in Ref.~\cite{Goeke:2005hb} and
consistent with the other articles; $(iii)$ the function $f_T^{\perp}$
here is different from Ref.~\cite{Goeke:2005hb}, where it was first
introduced, in order to maintain the symmetry with the other functions
and to have simpler expressions in the following results.  The
relation between the two notations is the following (recall that
$p_T^2 = - \bm{p}_T^2$):
\begin{align} 
f_T'\,\Bigr|_{\text{here}} &= \frac{p_T^2}{M^2} \,
f_T^{\perp \prime}\,\Bigr|_{\text{Ref.~\protect\cite{Goeke:2005hb}}}\, ,
&
f_T^{\perp}\,\Bigr|_{\text{here}} &= f_T^{\perp \prime}
-f_T^{\perp}\,\Bigr|_{\text{Ref.~\protect\cite{Goeke:2005hb}}}\, .
\end{align} 
The nomenclature of the distribution functions follows closely that of
Ref.~\cite{Mulders:1996dh}, sometimes referred to as ``Amsterdam
notation.''
We remark that a number of other notations exist for some of the
distribution functions, see e.g.\
Refs.~\cite{Ralston:1979ys,Barone:2001sp,Idilbi:2004vb}.  In particular, 
transverse-momentum-dependent functions at leading twist have been
widely discussed by Anselmino et al.~\cite{Anselmino:1996vq,
Anselmino:1996qx,Anselmino:2005sh}.  The connection between the
notation in these papers and the one used here is discussed in App. C
of Ref.~\cite{Anselmino:2005sh}.

We also list here the expressions for the traces of the correlator
$\Phi(x,p_T)$ 
from Ref.~\cite{Goeke:2005hb}. With $\Phi^{[\Gamma]} = \half \Tr [\slim
\Phi\, \Gamma\slim]$ we have
\begin{align} 
\label{e:1}
\Phi^{[\gamma^+]} & = 
 f_1
 - \frac{\eps_{T}^{\rho\sigma} p_{T \rho}^{}\slim S_{T \sigma}^{}}{M} \, 
   f_{1T}^{\perp}
 \,,
\\
\Phi^{[\gamma^+ \gamma_5]} & =
 S_L \, g_{1L}
 - \frac{{p}_{T} \cdott {S}_{T}}{M} \, 
   g_{1T}
 \,,
\\ 
\label{e:spi5}
\Phi^{[\ii \sigma^{\alpha +}\gamma_5]} & =
 S_{T}^{\alpha} \, h_{1}
 + S_L\,\frac{p_{T}^{\alpha}}{M} \, h_{1L}^{\perp}
\nonumber \\ & \quad
 - \frac{p_{T}^{\alpha}\slim p_{T}^{\rho}
     -\frac{1}{2}\,{p}_T^{2}\,g_T^{\alpha\rho}}{M^2}\, S_{T \rho} \, 
   h_{1T}^{\perp}
- \frac{\eps_{T}^{\alpha\rho} p_{T \rho}^{}}{M} \, 
   h_{1}^{\perp} 
 \,,
\\ \nonumber
\\
\Phi^{[1]} & =
 \frac{M}{P^+} \bigg[ e
 - \frac{\eps_{T}^{\rho\sigma} p_{T \rho}^{}\slim S_{T \sigma}^{}}{M} \, 
   e_{T}^{\perp}
 \bigg] ,
\\
\Phi^{[i \gamma_5]} & =
 \frac{M}{P^+} \bigg[ S_L \slim e_{L}
 - \frac{{p}_{T} \cdott {S}_{T}}{M} \, 
   e_{T}
 \bigg] ,
\\ 
\label{e:gi}
\Phi^{[\gamma^{\alpha}]} & =
 \frac{M}{P^+} \bigg[
 - \eps_{T}^{\alpha\rho} S_{T \rho}^{} \, 
   f_{T}
 - S_L \,\frac{\eps_{T}^{\alpha\rho} p_{T \rho}^{}}{M}\, 
   f_{L}^{\perp}
\nonumber \\ & \qquad
 -  \frac{p_T^{\alpha}\slim p_T^{\rho}
        -\frac{1}{2}\,{p}_T^{2}\,g_{T}^{\alpha\rho}}{M^2} 
        \,\eps^{}_{T \rho\sigma}\slim S_{T}^{\sigma}\, 
   f_{T}^{\perp}
 + \frac{p_{T}^{\alpha}}{M} f^{\perp}
 \bigg] ,
\\
\label{e:gi5}
\Phi^{[\gamma^{\alpha}\gamma_5]} & =
 \frac{M}{P^+} \bigg[
 S_{T}^{\alpha} \, g_{T}
 + S_L \, \frac{p_{T}^{\alpha}}{M}\, g_{L}^{\perp}
\nonumber \\ & \qquad
 - \frac{p_{T}^{\alpha}\slim p_{T}^{\rho}
     -\frac{1}{2}\,{p}_T^{2}\,g_T^{\alpha\rho}}{M^2} 
        \,S_{T \rho}^{}\,
   g_{T}^{\perp}
- \frac{\eps_{T}^{\alpha\rho} p_{T \rho}^{}}{M} \, 
  g^{\perp} 
 \bigg] ,
\\
\Phi^{[i\sigma^{\alpha\beta}\gamma_5]} & =
 \frac{M}{P^+} \bigg[ \frac{S_{T}^{\alpha}\slim p_{T}^{\beta} 
                     - p_{T}^{\alpha}\slim S_{T}^{\beta}}{M} \,
   h_{T}^{\perp}
 - \eps_{T}^{\alpha\beta} \, h
 \bigg] ,
\\
\Phi^{[i\sigma^{+-}\gamma_5]} & =
 \frac{M}{P^+} \bigg[ S_L \slim h_{L}
 - \frac{{p}_{T} \cdott {S}_{T}}{M} \, 
   h_{T}
 \bigg] ,
\end{align}
where $\alpha$ and $\beta$ are restricted to be transverse indices.
Here we made use of the combinations
\begin{align}
f_T^{}(x, p_T^2) &= 
  f_T'(x, p_T^2) - \frac{p_T^2}{2 M^2}\, f_T^{\perp}(x, p_T^2),
\\
g_T^{}(x, p_T^2) &= 
  g_T'(x, p_T^2) - \frac{p_T^2}{2 M^2}\, g_T^{\perp}(x, p_T^2),
\\
h_{1}(x, p_T^2) &= 
  h_{1T}(x, p_T^2) - \frac{p_T^2}{2 M^2}\, h_{1T}^{\perp}(x, p_T^2),
\end{align} 
to separate off terms that vanish upon integration of the correlator
over transverse momentum due to rotational symmetry.  The conversion
between the expression
in Eq.~\eqref{e:gi} and that in Eq.~(19) of Ref.~\cite{Goeke:2005hb} can be
carried out using the identity
\begin{equation}
 p_T^2\, \eps_{T}^{\alpha\rho} S_{T\rho}^{} = 
 p_T^\alpha\, \eps_{T}^{\rho\sigma} p_{T \rho}^{}\slim S_{T \sigma}^{}
 + (p_T \cdott S_T)\; \eps_{T}^{\alpha\rho} p_{T \rho}^{} \,,
\end{equation} 
which follows from the fact that there is no completely antisymmetric
tensor of rank three in two dimensions.

Integrating the correlator over the transverse momentum $p_T$ yields
\begin{align}
\Phi(x) = \int \de^2 \bm{p}_T\,\Phi(x,p_T) & = \frac{1}{2}\, \biggl\{ 
f_1\slim \nslash_+ 
+S_L\slim g_{1} \, \gamma_5\nslash_+
+h_{1}\,\frac{\bigl[\Sslash_T, \nslash_+ \bigr]\gamma_5}{2}
\biggr\} 
\nonumber \\ & \quad
+ \frac{M}{2 P^+}\,\biggl\{ 
e 
- \ii\, S_L\slim {e_{L}} \slim \gamma_5 
+{f_T}\,\epsilon_T^{\rho\sigma}S_{T \rho}\gamma_\sigma
+g_T\, \gamma_5\Sslash_T
\nonumber \\ & \quad \qquad \qquad 
+S_L\,h_{L} \,\frac{[\nslash_+, \nslash_-]\gamma_5}{2}
+ \ii \, {h} \frac{ \bigl[\nslash_+, \nslash_- \bigr]}{2} 
\biggr\} ,
\label{eq:phix}
\end{align}
where the functions on the r.h.s.\ depend only on $x$ and are given by
\begin{equation}
f_1(x) =  \int \de^2 \bm{p}_T\; f_{1}(x,p_T^2)
\end{equation} 
and so forth for the other functions.  We have retained here one common
exception of notation, namely
\begin{equation}
g_1(x) = \int \de^2 \bm{p}_T\; g_{1L}(x,p_T^2) .
\end{equation}   
Other notations are also in use for the leading-twist integrated
functions, in particular $f_1^q = q$ (unpolarized distribution
function), $g_{1}^q= \Delta q$ (helicity distribution function),
$h_1^q = \delta q = \Delta_T q$ (transversity distribution function).
The T-odd functions vanish due to time-reversal
invariance~\cite{Goeke:2005hb}
\begin{align}
\int \de^2 \bm{p}_T\; f_T(x,p_T^2)&=0,
&
\int \de^2 \bm{p}_T\; e_L(x,p_T^2) &=0,
&
\int \de^2 \bm{p}_T\; h(x,p_T^2) &=0. 
\label{e:tinvfT}
\end{align} 
We have kept them in Eq.~\eqref{eq:phix} so that one can readily obtain the
analogous fragmentation correlator, where such functions do not
necessarily vanish.

The fragmentation correlation function is defined as
\begin{multline}
\Delta_{ij}(z,k_T)  =\frac{1}{2z}\sum_X \, \int
  \frac{\de\xi^+  \de^2\bm{\xi}_T}{(2\pi)^{3}}\; e^{\ii k \cdot \xi}\,
    \langle 0|\, {\cal U}^{n_+}_{(+\infty,\xi)}
\,\psi_i(\xi)|h, X\rangle 
\langle h, X|
             \bar{\psi}_j(0)\,
{\cal U}^{n_+}_{(0,+\infty)}
|0\rangle \bigg|_{\xi^-=0}\,,
\\    
\label{e:delta}
\end{multline} 
with $k^- = P_h^-/z$ and the Wilson lines
\begin{align} 
{\cal U}^{n_+}_{(+\infty,\xi)} &\equiv 
{\cal U}^T{(\bm{\infty}_T, \bm{\xi}_T;+\infty^+)}\;
{\cal U}^{n_+}{(+\infty^+, \xi^+;\bm{\xi}_T)} ,
\\[0.2em]
{\cal U}^{n_+}_{(0,+\infty)} &\equiv
{\cal U}^{n_+}{(0^+, +\infty^+;\bm{0}_T)}\;
{\cal U}^T{(\bm{0}_T, \bm{\infty}_T; +\infty^+)} .
\end{align} 
The notation ${\cal U}^{n_+}{(a^+,b^+;\bm{c}_T)}$ indicates a Wilson
line running along the plus direction from $[0^-,a^+,\bm{c}_T]$ to
$[0^-,b^+,\bm{c}_T]$, while ${\cal U}^T{(a_T,b_T;\bm{c}^+)}$ indicates
a gauge link running in the transverse direction from
$[0^-,c^+,\bm{a}_T]$ to $[0^-,c^+,\bm{b}_T]$.  The definition written
above naturally applies for the correlation function appearing in
$e^+e^-$ annihilation.  For semi-inclusive DIS it seems more natural
to replace all occurrences of $+\infty^+$ in the gauge links by
$-\infty^+$ \cite{Boer:2003cm}.
However, in Ref.~\cite{Collins:2004nx} it was shown that factorization
can be derived in such a way that the fragmentation correlators in
both semi-inclusive DIS and $e^+e^-$ annihilation have gauge links
pointing to $+\infty^+$.

The fragmentation correlation function (for a spinless or an
unpolarized hadron) can be parameterized as
\begin{align} 
 \label{eq:delta} 
\Delta(z, k_T) &= 
\frac{1}{2} \, \biggl\{ D_1 \nslash_- + \ii H_1^\perp \frac{ \bigl[ 
  \kslash_T, \nslash_- \bigr]}{2M_h}\biggr\} 
\nonumber \\ &\quad
+ \frac{M_h}{2 P_h^-}\,\biggl\{ E +D^\perp \frac{\kslash_T}{M_h}+ \ii H 
\frac{ \bigl[\nslash_-, \nslash_+ \bigr]}{2} + G^\perp 
\g_5\,\frac{\eps_T^{\rho \sigma} \g_{\rho}\slim k_{T \sigma}}{M_h}\biggr\}, 
\end{align} 
where the functions on the r.h.s.\ depend on $z$ and $k_T^2$.  To be
complete, they should all carry also a flavor index, and the final
hadron type should be specified.  The correlation function $\Delta$
can be directly obtained from the correlation function $\Phi$ by
changing\footnote{%
The change of sign of the tensor $\eps_T$ is due to the exchange $n_+
  \leftrightarrow n_-$ in its definition \protect\eqref{e:eps-def}.}
\begin{align} 
n_+ &\leftrightarrow n_-,
&
\eps_T &\to -\eps_T,
&
P^+ &\to P_h^-, 
&
M & \to M_h,
&
x & \to 1/z,
\end{align} 
and replacing the distribution functions with the corresponding
fragmentation functions ($f$ is replaced with $D$ and all other
letters are capitalized).

The correlator integrated over transverse momentum reads
\begin{equation}
 \Delta(z) =  z^2 \int \de^2 \bm{k}_T\, \Delta(z, k_T) =
  D_1 \frac{\nslash_-}{2} + \frac{M_h}{2 P_h^-}\,\biggl\{ E +\ii H 
\frac{ \bigl[\nslash_-, \nslash_+ \bigr]}{2}\biggr\},
\end{equation} 
where the functions on the r.h.s.\ are defined as
\begin{equation}
 D_1 (z) = z^2 \int \de^2 \bm{k}_T\,D_1 (z, k_T^2)
\end{equation} 
and so forth for the other functions.  The prefactor $z^2$ appears
because $D_1(z,k_T)$ is a probability density w.r.t.\ the transverse
momentum $k'_T = -z k_T$ of the final-state hadron relative to the
fragmenting quark.  The fragmentation correlator for polarized
spin-half 
hadrons is parameterized in analogy to Eqs.~\eqref{eq:phi} and
\eqref{eq:phix}.  As already remarked, in this case the functions
$D_T$, $E_L$ and $H$ (the analogs of $f_T$, $e_L$ and $h$) do not
vanish because $|h, X\rangle$ is an interacting state that does not
transform into itself under time-reversal.  Of course, it should be
taken as an outgoing state in the fragmentation correlator.

\subsection{The quark-gluon-quark correlators}
\label{s:qgqcorr}

We now examine the quark-gluon-quark distribution correlation
functions~\cite{Boer:2003cm,Fetze}
\begin{align} 
\lf(\Phi_D^{\mu}
\rg)_{ij} (x,p_T)  
&= 
\int \frac{\de \xi^- \de^2 \bm{\xi}_T}{(2\pi)^{3}}\;
e^{\ii p \cdot \xi} \langle P| \bar{\psi}_j(0)\,
{\cal U}^{n_-}_{(0,+\infty)}\,
{\cal U}^{n_-}_{(+\infty,\xi)}\,
\ii D^{\mu}(\xi)\, 
\psi_i(\xi) |P \rangle 
\bigg|_{\xi^+=0} ,
\label{e:phiD}
\end{align}
which contain the covariant derivative $\ii D^{\mu}(\xi) = \ii
\partial^{\mu}+ g A^{\mu}$.  Using
$
{\cal U}^{n_-}_{(+\infty,\xi)}\, \ii D^{+}(\xi)\, \psi(\xi) =
  \ii \partial^+ \bigl[ {\cal U}^{n_-}_{(+\infty,\xi)}\, \psi(\xi) \bigr]
$
we can write
\begin{equation}
  \label{e:phiPlus}
\Phi_D^{+}(x,p_T) = x P^+ \Phi(x,p_T)
\end{equation}
for the plus-component of the correlator.  For the transverse
components we define a further correlator~\cite{Boer:2003cm}
\begin{align}
{\PhiA}^{\alpha} (x,p_T)  
&= 
\Phi_D^{\alpha} (x,p_T)
-  p_T^{\alpha}\, \Phi (x,p_T) ,
\label{e:phiA}
\end{align} 
which is manifestly gauge invariant.  It reduces to a correlator
defined as in Eq.~(\ref{e:phiD}) with the covariant derivative
$i D^{\mu}$ replaced by $g A_T^{\alpha}$ if one has ${\cal
U}^{n_-}_{(+\infty,\xi)} = 1$, which is the case in a light-cone gauge
$A^+=0$ with suitable boundary conditions at light-cone
infinity~\cite{Belitsky:2002sm}.  
    Notice that \eqref{e:phiD} does not have the most general form of a
    twist-three correlation function since it depends on the
    kinematics of one instead of two partons (i.e., on a single
    momentum fraction and a single transverse momentum).
    Correspondingly, the covariant derivative is taken at the same
    space-time point as one of the quark fields.  The tree-level
    result \eqref{eq1} for semi-inclusive DIS can be expressed in terms of
    only the quark-quark correlator \eqref{e:phi}, the quark-gluon-quark
    correlator \eqref{e:phiD}, and their fragmentation counterparts.  This is
    not too surprising since the kinematics of our process defines a
    single plus-momentum fraction $x$, a single minus-momentum
    fraction $z$, and a single transverse momentum $P_{h\perp}$.  No
    further momentum fraction can for instance be constructed from the
    ratio $P_{h\perp}^2 / Q^2$, which is replaced with zero in the
    kinematical limit we consider.

    The correlation function \eqref{e:phiA} can be decomposed
    as
\begin{align}
\lefteqn{\PhiA^{\alpha}(x,p_T) =}
\nonumber \\[0.2em]
 & \frac{x M}{2}\,
\biggl\{ 
\Bigl[
\bigl(\tilde{f}^\perp-\ii\slim \tilde{g}^{\perp} \bigr)
        \frac{p_{T \rho}^{}}{M} 
-\bigl(\tilde{f}_T'+ \ii\slim \tilde{g}_T'\bigr) 
     \,\epsilon_{T \rho\sigma}^{}\slim S_{T}^{\sigma}
-\bigl(\tilde{f}_s^{\perp}+\ii\,\tilde{g}_s^{\perp}\bigr)
     \frac{\epsilon_{T \rho \sigma}^{}\slim p_{T}^{\sigma}}{M} 
 \slim\Bigl]
\bigl(g_T^{\alpha \rho} - i \epsilon_T^{\alpha\rho} \gamma_5\bigr)
\nonumber \\[0.2em]
 & \quad\!
-\bigl(\tilde{h}_s + \ii\,\tilde{e}_s\bigr)
        \gamma_T^{\alpha}\,\gamma_5
+\Bigl[\bigl(\tilde{h} + \ii\,\tilde{e}\bigr)
  +\bigl( \tilde{h}_T^{\perp}
        - \ii\,\tilde{e}_T^{\perp}\bigr)\slim 
   \frac{\eps_T^{\rho \sigma} p_{T\rho}^{}\slim S_{T\sigma}^{}}{M} 
 \slim\Bigr]
  \ii \gamma_T^{\alpha}
+ \ldots \bigl(g_T^{\alpha \rho} 
               + \ii \epsilon_T^{\alpha\rho} \gamma_5\bigr)
\biggr\} \frac{\nslash_+}{2}\, ,
\nonumber \\
\label{eq:phideltag}
\end{align}
where the index $\alpha$ is restricted to be transverse here and in
the following equations.  The functions on the r.h.s.\ depend on $x$
and $p_T^2$, except for the functions with subscript $s$, which are
defined as in Eq.~(\ref{eq:shorthand}).  The last term in the curly
brackets is irrelevant for the construction of the hadronic tensor of
semi-inclusive DIS and has not been parameterized explicitly.  The
only relevant traces of the quark-gluon-quark correlator are
\begin{align}
\frac{1}{2M x} \Tr \big[\tilde{\Phi}_{A\slim \alpha}\,  
  \sigma^{\alpha+} \big] 
&= \tilde{h}  + i\,\tilde{e}
  + \frac{\eps_T^{\rho \sigma} p_{T\rho} S_{T\sigma}}{M}\,
    \bigl(\tilde{h}_T^{\perp}
  - i\,\tilde{e}_T^{\perp}\bigr),
\\ 
\frac{1}{2M x} \Tr  
\big[ \tilde{\Phi}_{A\slim \alpha}\,  
  \ii \sigma^{\alpha+} \gamma_5 \big] 
&=  S_L\,\bigl(\tilde{h}_L+ i\,\tilde{e}_L\bigr)
  - \frac{p_{T} \cdott S_{T}}{M}\,
  \bigl(\tilde{h}_T + i\,\tilde{e}_T\bigr),
\\
\frac{1}{2M x}\, 
  \Tr \big[\tilde{\Phi}_{A \rho}\slim (g_T^{\alpha\rho} 
          + i \epsilon_T^{\alpha\rho} \gamma_5)\slim \gamma^+ \big] &= 
\frac{p_T^\alpha}{M} 
        \bigl(\tilde{f}^\perp - i \tilde{g}^\perp\bigr)  
- \epsilon_T^{\alpha\rho}\slim S_{T\rho}^{}\,
        \bigl(\tilde{f}_T + i \tilde{g}_T\bigr)
\nonumber \\ &\quad 
\hspace{-3.5cm}
- S_L \,\frac{\epsilon_T^{\alpha\rho}\slim p_{T\rho}^{}}{M} \,
        \bigl(\tilde{f}_L^\perp + i\,\tilde{g}_L^\perp\bigr) 
-      \frac{p_T^{\alpha}\,p_T^{\rho}
     -\frac{1}{2}\,{p}_T^{2}\,g_{T}^{\alpha\rho}}{M^2}\, 
        \eps_{T \rho\sigma}^{}\slim S_{T}^{\sigma}
        \,  \bigl(\tilde{f}_T^{\perp} + i \tilde{g}_T^{\perp}\bigr) , 
\end{align}
where again we have used the combinations
\begin{align} 
\tilde{f}_T(x, p_T^2) &= \tilde{f}_T'(x, p_T^2)- \frac{p_T^2}{2 M^2}\,
\tilde{f}_T^{\perp}(x, p_T^2) ,
\\
\tilde{g}_T(x, p_T^2) &= \tilde{g}_T'(x, p_T^2)- \frac{p_T^2}{2 M^2}\,
\tilde{g}_T^{\perp}(x, p_T^2) .
\end{align} 
The above traces have been given already in
Ref.~\cite{Bacchetta:2004zf} for the terms without transverse
polarization, whereas the terms with transverse polarization were
partly discussed in Ref.~\cite{Mulders:1996dh} (the functions
$e_T^{\perp}$ and $f_T^{\perp}$ introduced in Ref.~\cite{Goeke:2005hb}
were missing).

Relations between correlation functions of different twist are provided
by the equation of motion for the quark field
\begin{equation}
  \label{e:quark-eom}
\bigl[\slim \ii \! \Dslash(\xi) - m \bigr]\slim \psi(\xi)
 = \bigl[ \gamma^+ \ii D^-(\xi) + \gamma^- \ii D^+(\xi) 
        + \gamma_T^\alpha\, \ii D_\alpha(\xi) - m \bigr]\slim \psi(\xi)
 = 0 ,
\end{equation}
where $m$ is the quark mass.  To make their general structure
transparent we decompose the correlators into terms of definite twist,
\begin{align}
\Phi &= \Phi_2 + \frac{M}{P^+}\, \Phi_3 
       + \left(\frac{M}{P^+}\right)^2 \Phi_4 ,
&
\frac{1}{M}\, \PhiA^{\alpha} &= \tilde{\Phi}^{\alpha}_{A,3}
  + \frac{M}{P^+}\, \tilde{\Phi}^{\alpha}_{A,4}
  + \left(\frac{M}{P^+}\right)^2 \tilde{\Phi}^{\alpha}_{A,5} ,
\end{align}
where the twist is indicated in the subscripts and the arguments
$(x,p_T)$ are suppressed for ease of writing.  One has ${\cal P}_+ \Phi_4 =
{\cal P}_- \Phi_2 = 0$ and ${\cal P}_+ \tilde{\Phi}^{\alpha}_{A,5} = {\cal P}_-
\tilde{\Phi}^{\alpha}_{A,3} = 0$, where ${\cal P}_+ = \half \gamma^-
\gamma^+$ and ${\cal P}_- = \half \gamma^+ \gamma^-$ are the projectors on
good and bad light-cone components, respectively~\cite{Jaffe:1996zw}. 
Projecting Eq.~\eqref{e:quark-eom} on its good components one obtains
for the correlators
\begin{multline}
  \label{e:eom-inter}
{\cal P}_+ \Bigl[\slim x M\, \gamma^- \Phi_3 
  + M \gamma_{T\rho}\slim \tilde{\Phi}^{\rho}_{A,3}
  + \pslash_T \slim \Phi_2 - m \Phi_2 \slim\Bigr]
\\
+ {\cal P}_+ \Bigl[\slim x M\, \gamma^- \Phi_4
  + M \gamma_{T\rho}^{}\slim \tilde{\Phi}^{\rho}_{A,4}
  + \pslash_T \slim \Phi_3 - m \Phi_3 \slim\Bigr] \frac{M}{P^+} 
= 0 ,
\end{multline}
where the term with $D^-$ has disappeared and the terms with $D^+$ and
$D^\alpha$ have been replaced using Eqs.~\eqref{e:phiPlus} and
\eqref{e:phiA}.  Multiplying this relation with one of the matrices
$\Gamma^+ = \{ \gamma^+, \gamma^+ \gamma_5, i\sigma^{\alpha+}\gamma_5 \}$,
which satisfy $\Gamma^+ {\cal P}_+ = \Gamma^+ (1- {\cal P}_-) = \Gamma^+$, and
taking the trace gives
\begin{equation}
  \label{e:eom-master}
\Tr \Gamma^+ \left[\slim \gamma^- x\slim \Phi_3
  + \gamma_{T\rho}^{}\slim \tilde{\Phi}^{\rho}_{A,3}
  + \frac{\pslash_T}{M} \slim \Phi_2 
  - \frac{m}{M}\slim \Phi_2 \slim\right]
= 0 ,
\end{equation}
where the terms multiplied by $M/P^+$ in Eq.~\eqref{e:eom-inter} have
disappeared because the trace of Dirac matrices cannot produce a term
that transforms like $P^+$ under boosts in the light-cone direction.
Inserting the parameterizations \eqref{eq:phi} and
\eqref{eq:phideltag} into \eqref{e:eom-master}, one finds the
following relations between T-even functions:
\begin{align}
x  e &=x  \tilde{e} + \frac{m}{M}\,f_1,
\phantom{\frac{m^2}{M}}
\label{e:etilde}
\\
x f^{\perp} &=x  \tilde{f}^{\perp}+ f_{1},
\phantom{\frac{m^2}{M}}
\label{eq:Cahn}
\\
x g_T' &= x \tilde{g}_T'+\frac{m}{M}\,h_{1T},
\phantom{\frac{m^2}{M}}
\\
x g_T^{\perp}&= x  \tilde{g}_T^{\perp}+ g_{1T}
 +\frac{m}{M}\,h_{1T}^{\perp},
\phantom{\frac{m^2}{M}}
\\
x g_T &= x \tilde{g}_T -\frac{p_T^2}{2 M^2}\,  g_{1T}+\frac{m}{
  M}\,h_{1},
\\
x g_L^{\perp} &= x  \tilde{g}_L^{\perp} + g_{1L}+\frac{m}{
  M}\,h_{1L}^{\perp}, 
\phantom{\frac{m^2}{M}}
\label{eq:CahnL}
\\
x h_L &= x \tilde{h}_L +\frac{p_T^2}{M^2}\, h_{1L}^{\perp} 
 +\frac{m}{M}\,g_{1L},
\\
x h_T &= x \tilde{h}_T - h_{1} +\frac{p_T^2}{2 M^2}\, h_{1T}^{\perp} 
 +\frac{m}{M}\,g_{1T},
\\
x  h_T^{\perp} &= x \tilde{h}_T^{\perp}+h_{1} 
 +\frac{p_T^2}{2 M^2}\, h_{1T}^{\perp}.
\end{align} 
These relations can be found in Ref.~\cite{Mulders:1996dh}, App.~C.
Neglecting quark-gluon-quark correlators 
(often referred to as the Wandzura-Wilczek
approximation) 
is equivalent to setting all functions with a tilde to
zero.
For T-odd functions we have the following relations:
\begin{align}
x e_L &= x  \tilde{e}_L,
\phantom{\frac{p_T^2}{M^2}}
\\
x e_T &= x  \tilde{e}_T,
\phantom{\frac{p_T^2}{M^2}}
\\
x e_T^{\perp} &= x  \tilde{e}_T^{\perp} + \frac{m}{M}\,f_{1T}^{\perp},
\phantom{\frac{p_T^2}{M^2}}
\label{eq:eTperptilde}
\\
x f_T'  &= x  \tilde{f}_T' + \frac{p_T^2}{M^2}\, f_{1T}^{\perp}, 
\\
x f_T^{\perp} &= x \tilde{f}_T^{\perp} +  f_{1T}^{\perp},
\phantom{\frac{p_T^2}{M^2}}
\label{eq:fTperptilde}
\\
x f_T &= x \tilde{f}_T +\frac{p_T^2}{2 M^2}\,  f_{1T}^{\perp},
\label{e:fT}
\\
x f_L^{\perp}&= x  \tilde{f}_L^{\perp},
\phantom{\frac{p_T^2}{M^2}}
\\
x g^{\perp} &=  x  \tilde{g}^{\perp}+\frac{m}{M}\,h_{1}^{\perp},
\phantom{\frac{p_T^2}{M^2}}
\label{eq:gperptilde}
\\
x h  &=x  \tilde{h} + \frac{p_T^2}{M^2}\, h_1^{\perp}.
\label{e:h}
\end{align} 
Most of these relations can be found in
Refs.~\cite{Boer:1998bw,Daniel}, and Eq.~(\ref{eq:gperptilde}) can be
inferred from Eq.~(13) in Ref.~\cite{Bacchetta:2004zf}.
Eqs.~(\ref{eq:eTperptilde}) and (\ref{eq:fTperptilde}) have not been
given before as they require the new functions introduced in
Ref.~\cite{Goeke:2005hb}.  
We emphasize that the constraints due to the
equations of motion remain valid in the presence of the appropriate
Wilson lines in the correlation functions.  All that is required for
the gauge link ${\cal U}_{(0,\xi)}$ between the quark fields is the
relation ${\cal U}_{(0,\xi)}\, \ii D^{+}(\xi) = \ii \partial^+ {\cal
U}_{(0,\xi)}$ leading to Eq.~\eqref{e:phiPlus}.  In contrast, the
so-called Lorentz invariance relations used in earlier work are
invalidated by the presence of the gauge links \cite{Goeke:2003az}.
We remark that if quark-gluon-quark correlators are neglected, 
the
time-reversal constraints (\ref{e:tinvfT}) require that $\int\! \de^2
p_T^{}\, p_T^2 \slim
f_{1T}^{\perp} (x,p_T^2) =0$ and $\int\! \de^2 p_T^{}\, p_T^2\slim
h_{1}^{\perp} (x,p_T^2) =0$.

The quark-gluon-quark fragmentation correlator analogous to
$\Phi_D$ is defined as
\begin{align} 
&\hspace{-0.4cm}\lf(\Delta_D^{\mu}
\rg)_{ij} (z,k_T) =
\nonumber \\
&\hspace{-0.4cm}\frac{1}{2z}\, \sum_X
\int \frac{\de \xi^+\, \de^2 \bm{\xi}_T}{(2\pi)^{3}}\;
e^{\ii k \cdot \xi}\,
\langle 0|\,
{\cal U}^{n_+}_{(+\infty,\xi)}\,
i D^{\mu}(\xi)\,
\,\psi_i(\xi)|h, X\rangle
\langle h, X|
             \bar{\psi}_j(0)\,
{\cal U}^{n_+}_{(0,+\infty)}
|0\rangle \bigg|_{\xi^-=0} .
\label{e:DeltaD}
\end{align} 
The transverse correlator
\begin{equation}
{\DeltaA}^{\alpha} (z,k_T) = 
\Delta_D^{\alpha} (z,k_T)
- k_T^{\alpha}\,\Delta (z,k_T)
\label{e:DeltaA}
\end{equation}
can be decomposed as
\begin{align} 
\DeltaA^{\alpha}(z,k_T) 
 = \frac{M_h}{2 z}\,
&\biggl\{ 
\bigl(\tilde{D}^\perp-\ii\, \tilde{G}^{\perp} \bigr)\slim  
   \frac{k_{T \rho}^{}}{M_h}\,
   \bigl(g_T^{\alpha \rho} + i \epsilon_T^{\alpha\rho} \gamma_5\bigr)
\nonumber \\ & \quad 
+ \bigl(\tilde{H} + \ii\,\tilde{E}\bigr)\, \ii \gamma_T^{\alpha}
+ \ldots
  \bigl(g_T^{\alpha \rho} - i \epsilon_T^{\alpha\rho} \gamma_5\bigr)
\biggr\} \frac{\nslash_-}{2} .
\label{eq:deltaA}
\end{align} 
Using the equation of motion for the quark field, 
the following relations can be established between the functions
appearing in the above correlator and the functions in the quark-quark
correlator~(\ref{eq:delta}):
\begin{align}
\frac{E}{z}  &=\frac{\tilde{E}}{z} +  \frac{m}{M_h}\,D_1,
\\
\frac{D^{\perp}}{z}  &=\frac{\tilde{D}^{\perp}}{z} +  D_1,
\\
\frac{G^{\perp}}{z}  &=\frac{\tilde{G}^{\perp}}{z} +
                \frac{m}{M_h}\,H_1^{\perp},
\\
\frac{H}{z}  &=\frac{\tilde{H}}{z} + \frac{k_T^2}{M_h^2}\, H_1^{\perp}.
\end{align}

\section{Results for structure functions}
\label{s:cross}

Inserting the parameterizations of the different correlators in the
expression \eqref{eq1} of the hadronic tensor and using the
equation-of-motion constraints just discussed, one can calculate the
leptoproduction cross section for semi-inclusive DIS and project out
the different structure functions appearing in
Eq.~(\ref{e:crossmaster}).  To have a compact notation for the results,
we introduce the unit vector $\hat{\bm{h}}=\bm{P}_{h \perp}/|\bm{P}_{h
  \perp}|$ and
the notation
\begin{equation}
{\cal C}\bigl[ w\slim f\slim D \bigr]
= \xbj\,
\sum_a e_a^2 \int d^2 \bm{p}_T\,  d^2 \bm{k}_T^{}
\, \delta^{(2)}\bigl(\bm{p}_T - \bm{k}_T^{} - \bm{P}_{h \perp}/z \bigr)
\,w(\bm{p}_T,\bm{k}_T^{})\,
f^a(\xbj,p_T^2)\,D^a(z,k_T^2) ,
\end{equation}
where $w(\bm{p}_T,\bm{k}_T^{})$ is an arbitrary function and the summation
runs over quarks and antiquarks.
The expressions for the structure functions appearing in
Eq.~(\ref{e:crossmaster}) are
\begin{align} 
\label{F_UUT}
\hspace{-3mm}
F_{UU ,T}
& 
= {\cal C}\bigl[ f_1 D_1 \bigr],
\phantom{\biggl[ \biggr]}
\\
\label{F_UUL}
\hspace{-3mm}
F_{UU ,L}
& 
= 0,
\phantom{\biggl[ \biggr]}
\\ 
\label{F_UUcosphi}
\hspace{-3mm}
F_{UU}^{\cos\phi_h}
& 
= \frac{2M}{Q}\,{\cal C}\biggl[
   - \frac{\hat{\bm{h}}\cdott \bm{k}_T^{}}{M_h}
   \biggl(\xbj  h\, H_{1}^{\perp } 
   + \frac{M_h}{M}\,f_1 \frac{\tilde{D}^{\perp }}{z}\biggr)
   - \frac{\hat{\bm{h}}\cdott \bm{p}_T}{M}
     \biggl(\xbj f^{\perp } D_1
   + \frac{M_h}{M}\,h_{1}^{\perp } \frac{\tilde{H}}{z}\biggr)\biggr],
\\
\label{F_UUcos2phi}
\hspace{-3mm}
F_{UU}^{\cos 2\phi_h}\!
& 
= {\cal C}\biggl[
   - \frac{2\, \bigl( \hat{\bm{h}}\cdott \bm{k}_T^{} \bigr)
   \,\bigl( \hat{\bm{h}}\cdott \bm{p}_T \bigr)
    -\bm{k}_T^{}\cdott \bm{p}_T}{M M_h}
    h_{1}^{\perp } H_{1}^{\perp }\biggr],
\\
\label{F_LUsinphi} 
\hspace{-3mm}
F_{LU}^{\sin\phi_h}
& 
=\frac{2M}{Q}\,{\cal C}\biggl[ 
- \frac{\hat{\bm{h}}\cdott \bm{k}_T^{}}{M_h}
   \biggl(\xbj  e \, H_1^{\perp } 
   +\frac{M_h}{M}\,f_1\frac{\tilde{G}^{\perp }}{z}\biggr)
   \!+\! \frac{\hat{\bm{h}}\cdott \bm{p}_T}{M}
   \biggl(\xbj  g^{\perp }  D_1 
   + \frac{M_h}{M}\, h_1^{\perp } \frac{\tilde{E}}{z} \biggr)\biggr],
\\
\label{F_ULsinphi} 
\hspace{-3mm}
F_{UL}^{\sin\phi_h}
&
 = \frac{2M}{Q}\,{\cal C}\biggl[
   - \frac{\hat{\bm{h}}\cdott \bm{k}_T^{}}{M_h}
    \biggl(\xbj  h_L  H_1^{\perp } 
   +\frac{M_h}{M}\,g_{1L}\frac{\tilde{G}^{\perp } }{z}\biggr)
   \!+\! \frac{\hat{\bm{h}}\cdott \bm{p}_T}{M}
    \biggl(\xbj f_{L}^{\perp }  D_1 
   - \frac{M_h}{M}\, h_{1L}^{\perp }  \frac{\tilde{H}}{z}\biggr)\biggr],\,
\\
\hspace{-3mm}
F_{UL}^{\sin 2\phi_h}\!
&
 = {\cal C}\biggl[
   -\frac{2\,\bigl( \hat{\bm{h}}\cdott \bm{k}_T^{} \bigr)
   \,\bigl( \hat{\bm{h}}\cdott \bm{p}_T \bigr)
   -\bm{k}_T^{}\cdott \bm{p}_T}{M M_h}
    h_{1L}^{\perp } H_{1}^{\perp }\biggr],
\\
\hspace{-3mm}
F_{LL}
&
 ={\cal C}\bigl[ g_{1L}  D_1\bigr], 
\phantom{\biggl[ \biggr]}
\\
\hspace{-3mm}
F_{LL}^{\cos \phi_h}
&
 =\frac{2M}{Q}\,{\cal C}\biggl[ \frac{\hat{\bm{h}}\cdott\bm{k}_T^{}}{M_h}
   \biggl(\xbj e_L  H_1^{\perp }
   - \frac{M_h}{M}\,g_{1L}   \frac{\tilde{D}^{\perp }}{z}\biggr)
   - \frac{\hat{\bm{h}}\cdott \bm{p}_T}{M}
   \biggl(\xbj  g_L^{\perp }   D_1
   +  \frac{M_h}{M}\,h_{1L}^{\perp } \frac{\tilde{E}}{z}\biggr)\biggr],
\end{align} 
\begin{align} 
F_{UT ,T}^{\sin\lf(\phi_h -\phi_S\rg)}
\!& 
={\cal C}\biggl[-\frac{\h\cdott\bm{p}_T}{M} f_{1T}^{\perp }  D_1\biggr],
\label{e:sivers}
\\
F_{UT ,L}^{\sin\lf(\phi_h -\phi_S\rg)}
\!&= 0,
\phantom{\biggl[ \biggr]}
\\
F_{UT}^{\sin\lf(\phi_h +\phi_S\rg)}
\!&
 ={\cal C}\biggl[-\frac{\h\cdott\bm{k}_T^{}}{M_h} h_{1} H_1^{\perp }\biggr],
\\
F_{UT}^{\sin\lf(3\phi_h -\phi_S\rg)}
\!&
 = 
   {\cal C}\biggl[
   \frac{2\, \bigl(\h\cdott \bm{p}_T \bigr)\, 
        \bigl( \bm{p}_T\cdott\bm{k}_T^{} \bigr)
   +\bm{p}_T^2\, \bigl(\h\cdott \bm{k}_T^{} \bigr)
   -4\, (\h\cdott\bm{p}_T)^2 \, (\h\cdott\bm{k}_T^{})}{2 M^2 M_h}
    \,h_{1T}^{\perp }   H_1^{\perp }\biggr],\,\,\,
\\
F_{UT}^{\sin \phi_S }
&
 = \frac{2M}{Q}\,{\cal C}\biggl\{
   \biggl(\xbj  f_T   D_1
   - \frac{M_h}{M} \, h_{1}  \frac{\tilde{H}}{z}\biggr) 
\nonumber \\ & \qquad
\hspace{-9mm}
   - \frac{\bm{k}_T^{}\cdott \bm{p}_T}{2 M M_h}\,
     \biggl[\biggl(\xbj  h_{T}  H_{1}^{\perp } 
   + \frac{M_h}{M} g_{1T} \,\frac{\tilde{G}^{\perp }}{z}\biggr)
   -  \biggl(\xbj  h_{T}^{\perp }  H_{1}^{\perp } 
   - \frac{M_h}{M} f_{1T}^{\perp } \,\frac{\tilde{D}^{\perp }}{z}
   \biggr) \biggr]\biggr\},
\\ 
F_{UT}^{\sin\lf(2\phi_h -\phi_S\rg)}
\!& 
= \frac{2M}{Q}\,{\cal C}\biggl\{
   \frac{2\, (\hat{\bm{h}}\cdott \bm{p}_T)^2 -\bm{p}_T^2}{2 M^2}\,
   \biggl(\xbj  f_T^{\perp }   D_1
   - \frac{M_h}{M} \, h_{1T}^{\perp }  \frac{\tilde{H}}{z}\biggr)
\nonumber \\ & \qquad
   - \frac{2\, \bigl( \hat{\bm{h}}\cdott \bm{k}_T^{} \bigr)
   \, \bigl( \hat{\bm{h}}\cdott \bm{p}_T \bigr)
   -\bm{k}_T^{}\cdott \bm{p}_T}{2 M M_h}\,
   \biggl[\biggl(\xbj  h_{T}  H_{1}^{\perp } 
   + \frac{M_h}{M} g_{1T} \,\frac{\tilde{G}^{\perp }}{z}\biggr)
\nonumber \\ & \qquad
\hspace{53mm}   
        +  \biggl(\xbj  h_{T}^{\perp }  H_{1}^{\perp } 
   - \frac{M_h}{M} f_{1T}^{\perp } \,\frac{\tilde{D}^{\perp }}{z}
   \biggr) \biggr]\biggr\},
\label{e:dUT}
\\
F_{LT}^{\cos(\phi_h -\phi_S)}
\!&
 ={\cal C}\biggl[ \frac{\h\cdott\bm{p}_T}{M} g_{1T}
D_1 \biggr] , 
\\
F_{LT}^{\cos \phi_S}
&
 = \frac{2M}{Q}\,{\cal C}\biggl\{
   -\biggl(\xbj  g_T   D_1
   + \frac{M_h}{M} \, h_{1}  \frac{\tilde{E}}{z}\biggr) 
\nonumber \\ & \qquad
\hspace{-9mm}   +\frac{\bm{k}_T^{}\cdott \bm{p}_T}{2 M M_h}\,
   \biggl[\biggl(\xbj  e_{T}  H_{1}^{\perp } 
   - \frac{M_h}{M} g_{1T} \,\frac{\tilde{D}^{\perp }}{z}\biggr)
   +  \biggl(\xbj  e_{T}^{\perp }  H_{1}^{\perp } 
   + \frac{M_h}{M} f_{1T}^{\perp } \,\frac{\tilde{G}^{\perp }}{z}
   \biggr) \biggr]\biggr\},
\\ 
F_{LT}^{\cos(2\phi_h - \phi_S)}
\!&
 = \frac{2M}{Q}\,{\cal C}\biggl\{
   -\frac{2\,(\hat{\bm{h}}\cdott \bm{p}_T)^2 -\bm{p}_T^2}{2 M^2}\,
   \biggl(\xbj  g_T^{\perp }   D_1
   + \frac{M_h}{M} \, h_{1T}^{\perp }  \frac{\tilde{E}}{z}\biggr)
\nonumber \\ & \qquad
   + \frac{2\, \bigl( \hat{\bm{h}}\cdott \bm{k}_T^{} \bigr)
   \, \bigl( \hat{\bm{h}}\cdott \bm{p}_T \bigr)
   -\bm{k}_T^{}\cdott \bm{p}_T}{2 M M_h}\,
   \biggl[\biggl(\xbj  e_{T}  H_{1}^{\perp } 
   - \frac{M_h}{M} g_{1T} \,\frac{\tilde{D}^{\perp }}{z}\biggr)
\nonumber \\ & \qquad
\hspace{53mm}   -  \biggl(\xbj  e_{T}^{\perp }  H_{1}^{\perp } 
   + \frac{M_h}{M} f_{1T}^{\perp } \,\frac{\tilde{G}^{\perp }}{z}
   \biggr) \biggr]\biggr\}.
\label{e:last-fun}
\end{align} 
Notice that distribution and fragmentation functions do not appear in
a symmetric fashion in these expressions: there are only twist-three
fragmentation functions with a tilde and only twist-three distribution
functions without tilde.  This asymmetry is not surprising because in
Eq.~(\ref{e:crossmaster}) the
structure functions themselves are introduced in an asymmetric way,
with azimuthal angles referring to the axis given by the four-momenta
of the target nucleon and the photon, rather than of the target nucleon
and the detected hadron.

Equations~\eqref{F_UUT} to \eqref{e:last-fun} are a main result of
this paper.  A few comments concerning the comparison with the
existing literature are in order here. First of all, it has to be
stressed that in much of the past literature a different
definition of the azimuthal angles has been used, whereas in the
present work we adhere to the Trento
conventions~\cite{Bacchetta:2004jz}. To compare with those papers, the
signs of $\phi_h$ and of $\phi_S$ have to be reversed.  The terms with
the distribution functions $f_T^\perp$ and $e_T^\perp$ have not been
given before.  All leading-twist structure functions here are
consistent with those given in Eqs.~(36) and (37) of
Ref.~\cite{Boer:1999uu} when only photon exchange is taken into
consideration.  The structure functions $F_{LU}^{\sin\phi_h}$ and
$F_{UL}^{\sin\phi_h}$ in our Eqs.~(\ref{F_LUsinphi}) and
(\ref{F_ULsinphi}) correspond to Eqs.~(16) and (25) in
Ref.~\cite{Bacchetta:2004zf}.  The other six twist-three structure
functions were partially given in the original work of Mulders and
Tangerman~\cite{Mulders:1996dh}, but excluding T-odd distribution
functions, assuming Gaussian transverse-momentum distributions, and
without the contributions from the fragmentation function~$G^{\perp}$.

The structure function $F_{UU}^{\cos\phi_h}$ is associated with the
so-called Cahn effect~\cite{Cahn:1978se,Cahn:1989yf}.  If one neglects 
the quark-gluon-quark
functions $\tilde{D}^\perp$ in
Eq.~(\ref{F_UUcosphi}) and $\tilde{f}^\perp$ in Eq.~(\ref{eq:Cahn}) as
well as the T-odd distribution functions $h$ and $h_1^\perp$, our result 
becomes
\begin{equation}
 F_{UU}^{\cos\phi_h}
\approx \frac{2M}{Q}\,{\cal C}\biggl[ -\frac{\hat{\bm{h}}\cdott
  \bm{p}_T}{M}\,f_1 D_1\biggr] .
\end{equation}   
This coincides with the $\cos\phi_h$ term calculated to order $1/Q$ 
in the parton
model with intrinsic transverse momentum included in distribution
and fragmentation functions, see e.g.\ Eqs.~(32) and (33) in
 Ref.~\cite{Anselmino:2005nn}.

Let us briefly mention some experimental results and phenomenological
analyses for the structure functions given above.  For simplicity we
do not distinguish between measurements of the structure functions and
of the associated spin or angular asymmetries, which correspond to the
ratio of the appropriate structure functions and $F_{UU,T} + \epsilon
F_{UU,L}$.
\begin{enumerate}
\item Measurements of the cross-section components containing the
  structure function $F_{UU}^{\cos\phi_h}$ have been reported in
  Refs.~\cite{Arneodo:1986cf,Adams:1993hs,Breitweg:2000qh,Chekanov:2006gt}.
  A description of the $\cos\phi_h$ modulation by the Cahn effect
  alone has been given in Ref.~\cite{Anselmino:2005nn}.  The
  same analysis can be applied to the structure function
  $F_{LL}^{\cos\phi_h}$, leading to the results of
  Ref.~\cite{Anselmino:2006yc}.
\item $F_{UU}^{\cos 2\phi_h}$ contains the functions $h_1^\perp$
  (Boer-Mulders function~\cite{Boer:1997nt}) and $H_1^\perp$ (Collins
  function~\cite{Collins:1993kk}). It has been measured in
  Refs.~\cite{Breitweg:2000qh,Chekanov:2006gt}.
\item The structure function $F_{LU}^{\sin\phi_h}$ has been recently
  measured by the CLAS collaboration~\cite{Avakian:2003pk}.
\item The structure function $F_{UL}^{\sin\phi_h}$ has been measured
  by HERMES~\cite{Airapetian:2005jc}.  The precise extraction of this
  observable requires care because in experiments the target is
  polarized along the direction of the lepton beam and not of the
  virtual photon~\cite{Korotkov:1999jx,Oganessyan:2002eq,Bacchetta:2004zf,Diehl:2005pc}.
  This implies that the longitudinal target-spin asymmetries measured
  in
  Refs.~\cite{Airapetian:2000tv,Airapetian:2001eg,Airapetian:2002mf}
  receive contributions not only from $F_{UL}^{\sin\phi_h}$, but at
  the same order in $1/Q$ also from $F_{UT,T}^{\sin\lf(\phi_h
  -\phi_S\rg)}$ and $F_{UT}^{\sin\lf(\phi_h +\phi_S\rg)}$ (see also
  the phenomenological studies of
  Refs.~\cite{Efremov:2001cz,Efremov:2001ia,Efremov:2002ut,Ma:2002ns,%
  Efremov:2003eq,Efremov:2003tf,Schweitzer:2003yr}).  In
  Ref.~\cite{Airapetian:2005jc} the HERMES collaboration has separated
  the different contributions to the experimental $\sin\phi_h$
  asymmetry with longitudinal target polarization and shown that
  $F_{UL}^{\sin\phi_h}$ is dominant in the kinematics of the
  measurement.
\item $F_{UT ,T}^{\sin\lf(\phi_h -\phi_S\rg)}$ contains the Sivers
  function~\cite{Sivers:1990cc} and has been recently measured for a
  proton target at HERMES~\cite{Airapetian:2004tw} and for a deuteron
  target at COMPASS~\cite{Alexakhin:2005iw}.  Extractions of the
  Sivers function from the experimental data were performed in
  Refs.~\cite{Anselmino:2005ea,Vogelsang:2005cs,Collins:2005ie} (see
  Ref.~\cite{Anselmino:2005an} for a comparison of the various
  extractions).
\item The structure function $F_{UT}^{\sin\lf(\phi_h +\phi_S\rg)}$
  contains the transversity distribution
  function~\cite{Ralston:1979ys,Barone:2001sp} and the Collins
  function.  As the previous structure function, it has been measured
  by HERMES~\cite{Airapetian:2004tw} on the proton and by
  COMPASS~\cite{Alexakhin:2005iw} on the deuteron.  Phenomenological
  studies have been presented in Ref.~\cite{Vogelsang:2005cs}, where
  information about the Collins function was extracted, and in
  Ref.~\cite{Efremov:2006qm}, where constraints on the transversity
  distribution function were obtained by using additional information
  from a Collins asymmetry measured in $e^+e^-$
  annihilation~\cite{Abe:2005zx}.
\end{enumerate}

Integration of Eqs.~(\ref{F_UUT}) to (\ref{e:last-fun}) 
over the transverse momentum $\bm{P}_{h\perp}$ of the
outgoing hadron leads to the following expressions for the integrated
structure functions in Eq.~(\ref{e:crossintsidis}):
\begin{align}
F_{UU ,T} &= \xbj\,\sum_a e_a^2\,f_1^a(\xbj)\,D_1^a(z),
\phantom{\biggl( \biggr)}
\\
F_{UU ,L} &= 0,
\phantom{\biggl( \sum_a \biggr)}
\\
F_{LL} &=\xbj\,\sum_a e_a^2\,g_1^a(\xbj)\,D_1^a(z),
\phantom{\biggl( \biggr)}
\\
F_{UT}^{\sin \phi_S}&=-\xbj\,\sum_a e_a^2\,
        \frac{2M_h}{Q}\,h_{1}^a(\xbj)\,\frac{\tilde{H}^a(z)}{z},
\label{e:FUTint}
\\
F_{LT}^{\cos \phi_S}&=-\xbj\,\sum_a e_a^2\, \frac{2M}{Q}\,
\biggl(\xbj  g_T^a(\xbj)   D_1^a(z)
   + \frac{M_h}{M} \, h_{1}^a(\xbj)\,  \frac{\tilde{E}^a(z)}{z}\biggr).
\end{align} 
Finally, the structure functions for totally inclusive DIS can be
obtained from Eqs.~(\ref{e:sidistodis}) and (\ref{dis-fun}) to
(\ref{dis-last}). 
This gives the standard results~\cite{Mulders:1996dh}
\begin{align}
\label{e:FT}
F_1 &= \frac{1}{2}\,\sum_a e_a^2\;f_1^a(\xbj),
\\
F_L &= 0,
\phantom{\sum_a}
\\
g_1 &= \frac{1}{2}\,\sum_a e_a^2\;g_1^a(\xbj),
\label{e:g1}
\\
\label{e:g12}
g_1 + g_2&= \frac{1}{2}\,\sum_a e_a^2\; g_T^a(\xbj),
\end{align} 
where given the accuracy of our calculation we have replaced $g_1 -
\gamma^2 g_2$ by $g_1$ in \eqref{e:g1}, and where we have used
\begin{align} 
\sum_h \int \de z \,z\,D_1^a(z) &=1,
&
\sum_h \int \de z \,\tilde{E}^a(z) &=0,
&
\sum_h \int \de z \,\tilde{H}^a(z) &=0.
\end{align}  
The first relation is the well-known momentum sum rule for fragmentation
functions. 
The second relation
 was already pointed out in Ref.~\cite{Jaffe:1993xb}. The sum
rule for $\tilde{H}$ follows from Eq.~(\ref{e:FUTint}) and the
time-reversal constraint (\ref{e:nosinphis}).  

In App.~\ref{a:jet}, we give results for one-jet production
at low transverse momentum in DIS.

\section{Conclusions}
\label{s:conc}

We have analyzed
one-particle inclusive deep inelastic
scattering off a polarized nucleon for low transverse momentum of the detected
hadron, starting 
from a general decomposition of the cross section in terms of 18
structure functions, given in Eq.~(\ref{e:crossmaster}) and expressed in a helicity basis
in App. \ref{a:sapeta}. 
Using tree-level factorization as discussed in Sec.~\ref{s:hadtens}, the
structure functions can be calculated up to subleading order in $1/Q$ (twist
three) using
transverse-momentum-dependent
quark-quark and quark-gluon-quark correlators. 
to subleading order, we need the parameterizations of the correlators
up to twist three, involving 24 parton distributions. In particular our
treatment includes those twist three
transverse momentum dependent functions that appear after the proper
inclusion of Wilson lines in the quark-quark correlators. We also give the
relations between the quark-quark
and quark-gluon-quark correlators that follow from the QCD equations of
motion.

Using the parameterization of the correlators we eventually expressed the
structure functions appearing in the cross section in terms of
transverse-momentum-dependent parton distribution and fragmentation
functions, see Eqs.~(\ref{F_UUT}) to (\ref{e:last-fun}). Several of
these results were already present in the literature, 
but never collected in a single paper. The
complete results for transversely polarized targets,
Eqs.~(\ref{e:sivers}) to (\ref{e:last-fun}), 
including all T-odd distribution functions, are
presented here for the first time.
It is straightforward to generalize the present results to include 
the production of a
polarized hadron.


\section*{Acknowledgments}

This research is part of the Integrated
Infrastructure Initiative ``Hadron Physics'' of the 
European Union under contract
number RII3-CT-2004-506078.
The work of M.D. is supported by the Helmholtz Association, contract
number VH-NG-004. 
The work of A.M. is partially 
supported by the Verbundforschung of the BMBF (06-Bo-103).

\appendix

\section{Structure functions in a helicity basis}
\label{a:sapeta}

Up to a kinematical factor, the structure functions we have introduced
in Eq.~\eqref{e:crossmaster} are simple combinations of cross sections
or interference terms for the subprocess $\gamma^* N\to h\slim X$ with
definite helicities of the nucleon and the virtual photon.  
Let us define
helicity structure functions
\begin{equation}
F^{ij}_{mn}(\xbj, Q^2, z, P_{h\perp}^2)
= \frac{Q^2 (1-x)}{4 \pi^3 \alpha}\,
\biggl( 1+\frac{\gamma^2}{2\xbj} \biggr)^{-1} 
  \,
  \frac{d\sigma^{ij}_{mn}}{dz\, d P_{h\perp}^2} ,
\end{equation}
in terms of the $\gamma^* p$ cross sections and interference terms
introduced in Ref.~\cite{Diehl:2005pc}, where $j$ and $i$ ($n$ and
$m$) are the helicities of the nucleon (virtual photon) in the
amplitude and its complex conjugate.  We then have
\begin{align}
F_{UU ,T} &=
  \frac{1}{2}\,\bigl(F^{++}_{++} + F^{--}_{++} \bigr) ,
&
F_{UU ,L} &=
  F^{++}_{00} ,
\nonumber \\
F_{UU}^{\cos\phi_h} &=
 -\frac{1}{\sqrt{2}}\,\text{Re}\,\bigl(F^{++}_{+0} + F^{--}_{+0} \bigr) ,
&
F_{UU}^{\cos 2\phi_h} &=
 -\text{Re}\,F^{++}_{+-} ,
\\ 
\nonumber \\
F_{LU}^{\sin\phi_h} &=
 -\frac{1}{\sqrt{2}}\,\text{Im}\,\bigl(F^{++}_{+0} + F^{--}_{+0} \bigr) ,
\\ 
\nonumber \\
F_{UL}^{\sin\phi_h} &=
 -\frac{1}{\sqrt{2}}\,\text{Im}\,\bigl(F^{++}_{+0} - F^{--}_{+0} \bigr) ,
&
F_{UL}^{\sin 2\phi_h} &=
 -\text{Im}\,F^{++}_{+-} ,
\\ 
\nonumber \\
F_{LL} &=
  \frac{1}{2}\,\bigl(F^{++}_{++} - F^{--}_{++} \bigr) ,
&
F_{LL}^{\cos \phi_h} &=
  -\frac{1}{\sqrt{2}}\,\text{Re}\, \bigl(F^{++}_{+0} - F^{--}_{+0} \bigr) ,
\\ 
\nonumber \\
F_{UT ,T}^{\sin\lf(\phi_h -\phi_S\rg)} &=
 -\text{Im}\,F^{+-}_{++} ,
&
F_{UT ,L}^{\sin\lf(\phi_h -\phi_S\rg)} &=
 -\text{Im}\,F^{+-}_{00} , \phantom{\frac{1}{\sqrt{2}}}
\nonumber \\
F_{UT}^{\sin\lf(\phi_h +\phi_S\rg)} &=
 -\frac{1}{2}\,\text{Im}\,F^{+-}_{+-} ,
&
F_{UT}^{\sin\lf(3\phi_h -\phi_S\rg)} &=
 -\frac{1}{2}\,\text{Im}\,F^{-+}_{+-} ,
\nonumber \\
F_{UT}^{\sin \phi_S } &=
 -\frac{1}{\sqrt{2}}\,\text{Im}\,F^{+-}_{+0} ,
&
F_{UT}^{\sin\lf(2\phi_h -\phi_S\rg)} &=
 -\frac{1}{\sqrt{2}}\,\text{Im}\,F^{-+}_{+0} ,
\\ 
\nonumber \\
F_{LT}^{\cos(\phi_h -\phi_S)} &=
  \text{Re}\,F^{+-}_{++} ,
&
F_{LT}^{\cos \phi_S}  &=
 -\frac{1}{\sqrt{2}}\,\text{Re}\,F^{+-}_{+0} ,
\nonumber \\
F_{LT}^{\cos(2\phi_h - \phi_S)} &=
 -\frac{1}{\sqrt{2}}\,\text{Re}\,F^{-+}_{+0} .
\end{align}
Comparing with our results~\eqref{F_UUT} to \eqref{e:last-fun} we find
a number of simple patterns.  The leading structure functions in the
$1/Q$ expansion are those where the photon is transverse in both the
amplitude and its conjugate, and the subleading structure functions
correspond to the interference between transverse and longitudinal
photon polarizations.  The two structure functions $F_{UU,L}^{}$ and
$F_{UT ,L}^{\sin\lf(\phi_h -\phi_S\rg)}$ involve only longitudinally
polarized photons.  In the kinematics we consider, they are of order
$1/Q^2$ and thus beyond the accuracy to which we have calculated.  We
finally remark that the number of transverse momentum factors
appearing in the different structure functions~\eqref{F_UUT} to
\eqref{e:last-fun} can be related to the mismatch between the helicity
differences $(m-i)$ and $(n-j)$ using angular momentum conservation,
see Ref.~\cite{Diehl:2005pc}.

\section{Semi-inclusive jet production}
\label{a:jet}

In this appendix, we take into consideration the process 
\begin{equation}
  \label{jetsidis}
\ell(l) + N(P) \to \ell(l') + \text{jet}(P_j) + X
\end{equation}
in the kinematical limit of large $Q^2$ at fixed $x$ and
$P_{h\perp}^2$. In the context of our tree-level calculation, we identify the
jet with the quark scattered from the virtual photon. We then have $z=1$ and   
the cross section formula is identical to
Eq.~(\ref{e:crossmaster}), except that it
is not differential in $z$.  
Correspondingly, the structure functions do not depend
on this variable.  The structure functions for the
process~\eqref{jetsidis} can be obtained from those of one-particle
inclusive DIS in Eqs.~(\ref{F_UUT}) to \eqref{e:last-fun} by replacing
$D_1(z,k_T^2)$ with $\delta(1-z)\slim \delta^{(2)}(\bm{k}_T)$, setting
all other 
fragmentation functions to zero and integrating over $z$.  This gives
\begin{align}
F_{UU ,T } &= \xbj\,\sum_a e_a^2\,
        f_1^a(x,P_{j\perp}^2)
        ,
\\
F_{UU}^{\cos\phi_h} &= - \xbj\,\sum_a e_a^2\,
        \frac{2|\bm{P}_{j \perp}|}{Q}\,\xbj  f^{\perp a}(x,P_{j\perp}^2)
        ,
\\
F_{LU}^{\sin\phi_h} &= \xbj\,\sum_a e_a^2\,
        \frac{2|\bm{P}_{j \perp}|}{Q}\,\xbj  g^{\perp a}(x,P_{j\perp}^2)
        ,
\\
F_{UL}^{\sin\phi_h} &= \xbj\,\sum_a e_a^2\,
        \frac{2|\bm{P}_{j \perp}|}{Q}\,\xbj  f_{L}^{\perp a}(x,P_{j\perp}^2)
        ,
\\
F_{LL} &= \xbj\,\sum_a e_a^2\,
        g_{1L}^a(x,P_{j\perp}^2)
        ,
\\
F_{LL}^{\cos \phi_h} &= -\xbj\,\sum_a e_a^2\,
        \frac{2|\bm{P}_{j \perp}|}{Q}\,\xbj  g_{L}^{\perp a}(x,P_{j\perp}^2)
        ,
\\
F_{UT ,T}^{\sin\lf(\phi_h -\phi_S\rg)} &= -\xbj\,\sum_a e_a^2\,
        \frac{|\bm{P}_{j \perp}|}{M}\, f_{1T}^{\perp a}(x,P_{j\perp}^2)
        ,
\\
F_{UT}^{\sin \phi_S} &= \xbj\,\sum_a e_a^2\,
        \frac{2M}{Q}\,\xbj  f_{T}^{a}(x,P_{j\perp}^2)
        ,
\label{e:sinphisjet}
\\
F_{UT}^{\sin\lf(2\phi_h -\phi_S\rg)} &= \xbj\,\sum_a e_a^2\,
        \frac{|\bm{P}_{j \perp}|^2}{M Q}\,
                        \,\xbj  f_{T}^{\perp a}(x,P_{j\perp}^2)
        ,
\\
F_{LT}^{\cos(\phi_h -\phi_S)} &= \xbj\,\sum_a e_a^2\,
        \frac{|\bm{P}_{j \perp}|}{M}\, g_{1T}^{a}(x,P_{j\perp}^2)
        ,
\\
F_{LT}^{\cos \phi_S} &= -\xbj\,\sum_a e_a^2\,
        \frac{2M}{Q}\,\xbj  g_{T}^{a}(x,P_{j\perp}^2)
        ,
\\
F_{LT}^{\cos(2\phi_h - \phi_S)} &= -\xbj\,\sum_a e_a^2\,
        \frac{|\bm{P}_{j \perp}|^2}{M Q}\,
                        \,\xbj  g_{T}^{\perp a}(x,P_{j\perp}^2)
        ,
\end{align} 
whereas the remaining 6 structure functions are zero. 
The results for the terms with indices $UU$, $LL$, and $LT$ correspond to
those in Ref.~\cite{Mulders:1996dh}, Eqs.~(119) to (121). 
 The results for
the terms with indices $LU$ and $UL$ correspond to those in
Ref.~\cite{Bacchetta:2004zf}.
Integration of the cross section 
over $\bm{P}_{h\perp}$ leads to the results for inclusive
DIS in Eqs.~(\ref{e:FT}) to \eqref{e:g12}.  Most terms vanish due to
the angular integration, and $F_{UT}^{\sin \phi_S}$ in
Eq.~(\ref{e:sinphisjet}) vanishes due to the
time-reversal condition~(\ref{e:tinvfT}).

\bibliography{mybiblio}

\end{document}